
%
%
%
\def\unredoffs{} \def\redoffs{\voffset=-.31truein\hoffset=-.59truein}
\def\speclscape{\special{ps: landscape}}
%
%
%
%
\newbox\leftpage \newdimen\fullhsize \newdimen\hstitle \newdimen\hsbody
\tolerance=1000\hfuzz=2pt
\catcode`\@=11 
\def\bigans{b }
\def\answ{b }
\ifx\answ\bigans\message{(This will come out unreduced.}
\magnification=1200\unredoffs\baselineskip=16pt plus 2pt minus 1pt
\hsbody=\hsize \hstitle=\hsize 
\else\message{(This will be reduced.} \let\l@r=L
\magnification=1000\baselineskip=16pt plus 2pt minus 1pt \vsize=7truein
\redoffs \hstitle=8truein\hsbody=4.75truein\fullhsize=10truein\hsize=\hsbody
\output={\ifnum\pageno=0 
  \shipout\vbox{\speclscape{\hsize\fullhsize\makeheadline}
    \hbox to \fullhsize{\hfill\pagebody\hfill}}\advancepageno
  \else
  \almostshipout{\leftline{\vbox{\pagebody\makefootline}}}\advancepageno
  \fi}
\def\almostshipout#1{\if L\l@r \count1=1 \message{[\the\count0.\the\count1]}
      \global\setbox\leftpage=#1 \global\let\l@r=R
 \else \count1=2
  \shipout\vbox{\speclscape{\hsize\fullhsize\makeheadline}
      \hbox to\fullhsize{\box\leftpage\hfil#1}}  \global\let\l@r=L\fi}
\fi
%
\newcount\yearltd\yearltd=\year\advance\yearltd by -1900

\def\Title#1#2{\nopagenumbers\abstractfont\hsize=\hstitle\rightline{#1}%
\vskip 1in\centerline{\titlefont #2}\abstractfont\vskip .5in\pageno=0}
\def\Date#1{\vfill\leftline{#1}\tenpoint\supereject\global\hsize=\hsbody%
\footline={\hss\tenrm\folio\hss}}
%

\def\draftmode{\message{ DRAFTMODE }\def\draftdate{{\rm preliminary draft:
\number\month/\number\day/\number\yearltd\ \ \hourmin}}%
\headline={\hfil\draftdate}\writelabels\baselineskip=20pt plus 2pt minus 2pt
 {\count255=\time\divide\count255 by 60 \xdef\hourmin{\number\count255}
  \multiply\count255 by-60\advance\count255 by\time
  \xdef\hourmin{\hourmin:\ifnum\count255<10 0\fi\the\count255}}}
\def\nolabels{\def\wrlabeL##1{}\def\eqlabeL##1{}\def\reflabeL##1{}}
\def\writelabels{\def\wrlabeL##1{\leavevmode\vadjust{\rlap{\smash%
{\line{{\escapechar=` \hfill\rlap{\sevenrm\hskip.03in\string##1}}}}}}}%
\def\eqlabeL##1{{\escapechar-1\rlap{\sevenrm\hskip.05in\string##1}}}%
\def\reflabeL##1{\noexpand\llap{\noexpand\sevenrm\string\string\string##1}}}
\nolabels
%
\global\newcount\secno \global\secno=0
\global\newcount\meqno \global\meqno=1
\def\newsec#1{\global\advance\secno by1\message{(\the\secno. #1)}
\global\subsecno=0\eqnres@t\noindent{\bf\the\secno. #1}
\writetoca{{\secsym} {#1}}\par\nobreak\medskip\nobreak}
\def\eqnres@t{\xdef\secsym{\the\secno.}\global\meqno=1\bigbreak\bigskip}
\def\sequentialequations{\def\eqnres@t{\bigbreak}}\xdef\secsym{}
\global\newcount\subsecno \global\subsecno=0
\def\subsec#1{\global\advance\subsecno by1\message{(\secsym\the\subsecno.
#1)}
\ifnum\lastpenalty>9000\else\bigbreak\fi
\noindent{\it\secsym\the\subsecno. #1}\writetoca{\string\quad
{\secsym\the\subsecno.} {#1}}\par\nobreak\medskip\nobreak}
\def\appendix#1#2{\global\meqno=1\global\subsecno=0\xdef\secsym{\hbox{#1.}}
\bigbreak\bigskip\noindent{\bf Appendix #1. #2}\message{(#1. #2)}
\writetoca{Appendix {#1.} {#2}}\par\nobreak\medskip\nobreak}
%
%
\def\eqnn#1{\xdef #1{(\secsym\the\meqno)}\writedef{#1\leftbracket#1}%
\global\advance\meqno by1\wrlabeL#1}
\def\eqna#1{\xdef #1##1{\hbox{$(\secsym\the\meqno##1)$}}
\writedef{#1\numbersign1\leftbracket#1{\numbersign1}}%
\global\advance\meqno by1\wrlabeL{#1$\{\}$}}
\def\eqn#1#2{\xdef #1{(\secsym\the\meqno)}\writedef{#1\leftbracket#1}%
\global\advance\meqno by1$$#2\eqno#1\eqlabeL#1$$}
%
\newskip\footskip\footskip14pt plus 1pt minus 1pt 
\def\footnotefont{\ninepoint}\def\f@t#1{\footnotefont #1\@foot}
\def\f@@t{\baselineskip\footskip\bgroup\footnotefont\aftergroup\@foot\let\next}
\setbox\strutbox=\hbox{\vrule height9.5pt depth4.5pt width0pt}
\global\newcount\ftno \global\ftno=0
\def\foot{\global\advance\ftno by1\footnote{$^{\the\ftno}$}}
%
\newwrite\ftfile
\def\footend{\def\foot{\global\advance\ftno by1\chardef\wfile=\ftfile
$^{\the\ftno}$\ifnum\ftno=1\immediate\openout\ftfile=foots.tmp\fi%
\immediate\write\ftfile{\noexpand\smallskip%
\noexpand\item{f\the\ftno:\ }\pctsign}\findarg}%
\def\footatend{\vfill\eject\immediate\closeout\ftfile{\parindent=20pt
\centerline{\bf Footnotes}\nobreak\bigskip\input foots.tmp }}}
\def\footatend{}
%
%
\global\newcount\refno \global\refno=1
\newwrite\rfile
\def\ref{[\the\refno]\nref}
\def\nref#1{\xdef#1{[\the\refno]}\writedef{#1\leftbracket#1}%
\ifnum\refno=1\immediate\openout\rfile=refs.tmp\fi
\global\advance\refno by1\chardef\wfile=\rfile\immediate
\write\rfile{\noexpand\item{#1\ }\reflabeL{#1\hskip.31in}\pctsign}\findarg}
\def\findarg#1#{\begingroup\obeylines\newlinechar=`\^^M\pass@rg}
{\obeylines\gdef\pass@rg#1{\writ@line\relax #1^^M\hbox{}^^M}%
\gdef\writ@line#1^^M{\expandafter\toks0\expandafter{\striprel@x #1}%
\edef\next{\the\toks0}\ifx\next\em@rk\let\next=\endgroup\else\ifx\next\empty%
\else\immediate\write\wfile{\the\toks0}\fi\let\next=\writ@line\fi\next\relax}}
\def\striprel@x#1{} \def\em@rk{\hbox{}}
\def\lref{\begingroup\obeylines\lr@f}
\def\lr@f#1#2{\gdef#1{\ref#1{#2}}\endgroup\unskip}
\def\semi{;\hfil\break}
\def\addref#1{\immediate\write\rfile{\noexpand\item{}#1}} 
\def\startrefs#1{\immediate\openout\rfile=refs.tmp\refno=#1}
\def\xref{\expandafter\xr@f}\def\xr@f[#1]{#1}
\def\refs#1{\count255=1[\r@fs #1{\hbox{}}]}
\def\r@fs#1{\ifx\und@fined#1\message{reflabel \string#1 is undefined.}%
\nref#1{need to supply reference \string#1.}\fi%
\vphantom{\hphantom{#1}}\edef\next{#1}\ifx\next\em@rk\def\next{}%
\else\ifx\next#1\ifodd\count255\relax\xref#1\count255=0\fi%
\else#1\count255=1\fi\let\next=\r@fs\fi\next}
%

%
\newwrite\ffile\global\newcount\figno \global\figno=1
\def\fig{fig.~\the\figno\nfig}
\def\nfig#1{\xdef#1{fig.~\the\figno}%
\writedef{#1\leftbracket fig.\noexpand~\the\figno}%
\ifnum\figno=1\immediate\openout\ffile=figs.tmp\fi\chardef\wfile=\ffile%
\immediate\write\ffile{\noexpand\medskip\noexpand\item{Fig.\ \the\figno. }
\reflabeL{#1\hskip.55in}\pctsign}\global\advance\figno by1\findarg}
\def\xfig{\expandafter\xf@g}\def\xf@g fig.\penalty\@M\ {}
\def\figs#1{figs.~\f@gs #1{\hbox{}}}
\def\f@gs#1{\edef\next{#1}\ifx\next\em@rk\def\next{}\else
\ifx\next#1\xfig #1\else#1\fi\let\next=\f@gs\fi\next}
\newwrite\lfile
{\escapechar-1\xdef\pctsign{\string\%}\xdef\leftbracket{\string\{}
\xdef\rightbracket{\string\}}\xdef\numbersign{\string\#}}

\def\writestop{\def\writestoppt{\immediate\write\lfile{\string\pageno%
\the\pageno\string\startrefs\leftbracket\the\refno\rightbracket%
\string\def\string\secsym\leftbracket\secsym\rightbracket%
\string\secno\the\secno\string\meqno\the\meqno}\immediate\closeout\lfile}}
\def\writestoppt{}\def\writedef#1{}
\def\seclab#1{\xdef #1{\the\secno}\writedef{#1\leftbracket#1}\wrlabeL{#1=#1}}
\def\subseclab#1{\xdef #1{\secsym\the\subsecno}%
\writedef{#1\leftbracket#1}\wrlabeL{#1=#1}}
\newwrite\tfile \def\writetoca#1{}
\def\leaderfill{\leaders\hbox to 1em{\hss.\hss}\hfill}
\def\writetoc{\immediate\openout\tfile=toc.tmp
   \def\writetoca##1{{\edef\next{\write\tfile{\noindent ##1
   \string\leaderfill {\noexpand\number\pageno} \par}}\next}}}

%
\catcode`\@=12 
%
\edef\tfontsize{\ifx\answ\bigans scaled\magstep3\else scaled\magstep4\fi}
\font\titlerm=cmr10 \tfontsize \font\titlerms=cmr7 \tfontsize
\font\titlermss=cmr5 \tfontsize \font\titlei=cmmi10 \tfontsize
\font\titleis=cmmi7 \tfontsize \font\titleiss=cmmi5 \tfontsize
\font\titlesy=cmsy10 \tfontsize \font\titlesys=cmsy7 \tfontsize
\font\titlesyss=cmsy5 \tfontsize \font\titleit=cmti10 \tfontsize
\skewchar\titlei='177 \skewchar\titleis='177 \skewchar\titleiss='177
\skewchar\titlesy='60 \skewchar\titlesys='60 \skewchar\titlesyss='60
\def\titlefont{\def\rm{\fam0\titlerm}
\textfont0=\titlerm \scriptfont0=\titlerms \scriptscriptfont0=\titlermss
\textfont1=\titlei \scriptfont1=\titleis \scriptscriptfont1=\titleiss
\textfont2=\titlesy \scriptfont2=\titlesys \scriptscriptfont2=\titlesyss
\textfont\itfam=\titleit \def\it{\fam\itfam\titleit}\rm}
 \ifx\answ\bigans\else scaled\magstep1\fi
\ifx\answ\bigans\def\abstractfont{\tenpoint}\else
\font\abssl=cmsl10 scaled \magstep1
\font\absrm=cmr10 scaled\magstep1 \font\absrms=cmr7 scaled\magstep1
\font\absrmss=cmr5 scaled\magstep1 \font\absi=cmmi10 scaled\magstep1
\font\absis=cmmi7 scaled\magstep1 \font\absiss=cmmi5 scaled\magstep1
\font\abssy=cmsy10 scaled\magstep1 \font\abssys=cmsy7 scaled\magstep1
\font\abssyss=cmsy5 scaled\magstep1 \font\absbf=cmbx10 scaled\magstep1
\skewchar\absi='177 \skewchar\absis='177 \skewchar\absiss='177
\skewchar\abssy='60 \skewchar\abssys='60 \skewchar\abssyss='60
\def\abstractfont{\def\rm{\fam0\absrm}
\textfont0=\absrm \scriptfont0=\absrms \scriptscriptfont0=\absrmss
\textfont1=\absi \scriptfont1=\absis \scriptscriptfont1=\absiss
\textfont2=\abssy \scriptfont2=\abssys \scriptscriptfont2=\abssyss
\textfont\itfam=\bigit \def\it{\fam\itfam\bigit}\def\footnotefont{\tenpoint}%
\textfont\slfam=\abssl \def\sl{\fam\slfam\abssl}%
\textfont\bffam=\absbf \def\bf{\fam\bffam\absbf}\rm}\fi
\def\tenpoint{\def\rm{\fam0\tenrm}
\textfont0=\tenrm \scriptfont0=\sevenrm \scriptscriptfont0=\fiverm
\textfont1=\teni  \scriptfont1=\seveni  \scriptscriptfont1=\fivei
\textfont2=\tensy \scriptfont2=\sevensy \scriptscriptfont2=\fivesy
\textfont\itfam=\tenit
\def\it{\fam\itfam\tenit}\def\footnotefont{\ninepoint}%
\textfont\bffam=\tenbf \def\bf{\fam\bffam\tenbf}\def\sl{\fam\slfam\tensl}\rm}
\font\ninerm=cmr9 \font\sixrm=cmr6 \font\ninei=cmmi9 \font\sixi=cmmi6
\font\ninesy=cmsy9 \font\sixsy=cmsy6 \font\ninebf=cmbx9
\font\nineit=cmti9 \font\ninesl=cmsl9 \skewchar\ninei='177
\skewchar\sixi='177 \skewchar\ninesy='60 \skewchar\sixsy='60
\def\ninepoint{\def\rm{\fam0\ninerm}
\textfont0=\ninerm \scriptfont0=\sixrm \scriptscriptfont0=\fiverm
\textfont1=\ninei \scriptfont1=\sixi \scriptscriptfont1=\fivei
\textfont2=\ninesy \scriptfont2=\sixsy \scriptscriptfont2=\fivesy
\textfont\itfam=\ninei \def\it{\fam\itfam\nineit}\def\sl{\fam\slfam\ninesl}%
\textfont\bffam=\ninebf \def\bf{\fam\bffam\ninebf}\rm}
%
%

\hyphenation{anom-aly anom-alies coun-ter-term coun-ter-terms}
\def\inv{^{\raise.15ex\hbox{${\scriptscriptstyle -}$}\kern-.05em 1}}

\def\Dsl{\,\raise.15ex\hbox{/}\mkern-13.5mu D} 
\def\dsl{\raise.15ex\hbox{/}\kern-.57em\partial}

\font\bigit=cmti10 scaled \magstep1
\def\lspace{\ifx\answ\bigans{}\else\qquad\fi}
\def\lbspace{\ifx\answ\bigans{}\else\hskip-.2in\fi} 
\def\boxeqn#1{\vcenter{\vbox{\hrule\hbox{\vrule\kern3pt\vbox{\kern3pt
           \hbox{${\displaystyle #1}$}\kern3pt}\kern3pt\vrule}\hrule}}}
\def\mbox#1#2{\vcenter{\hrule \hbox{\vrule height#2in
               \kern#1in \vrule} \hrule}}  
%

\def\e#1{{\rm e}^{^{\textstyle#1}}}

\def\darr#1{\raise1.5ex\hbox{$\leftrightarrow$}\mkern-16.5mu #1}

\def\roughly#1{\raise.3ex\hbox{$#1$\kern-.75em\lower1ex\hbox{$\sim$}}}



\def\IB{\relax\hbox{$\inbar\kern-.3em{\rm B}$}}
\def\IC{\relax\hbox{$\inbar\kern-.3em{\rm C}$}}
\def\ID{\relax\hbox{$\inbar\kern-.3em{\rm D}$}}
\def\IE{\relax\hbox{$\inbar\kern-.3em{\rm E}$}}
\def\IF{\relax\hbox{$\inbar\kern-.3em{\rm F}$}}
\def\IG{\relax\hbox{$\inbar\kern-.3em{\rm G}$}}
\def\IGa{\relax\hbox{${\rm I}\kern-.18em\Gamma$}}
\def\IH{\relax{\rm I\kern-.18em H}}
\def\IK{\relax{\rm I\kern-.18em K}}
\def\II{\relax{\rm I\kern-.18em I}}
\def\IL{\relax{\rm I\kern-.18em L}}
\def\IP{\relax{\rm I\kern-.18em P}}
\def\IR{\relax{\rm I\kern-.18em R}}
\def\IZ{\relax\ifmmode\mathchoice {\hbox{\cmss Z\kern-.4em Z}}{\hbox{\cmss
Z\kern-.4em Z}} {\lower.9pt\hbox{\cmsss Z\kern-.4em Z}}
{\lower1.2pt\hbox{\cmsss Z\kern-.4em Z}}\else{\cmss Z\kern-.4em Z}\fi}

\def\IB{\relax{\rm I\kern-.18em B}}
\def\IC{{\relax\hbox{$\inbar\kern-.3em{\rm C}$}}}
\def\ID{\relax{\rm I\kern-.18em D}}
\def\IE{\relax{\rm I\kern-.18em E}}
\def\IF{\relax{\rm I\kern-.18em F}}


\def\CW {{\cal W}}

\def\p{\partial}
\def\pa{\partial}
\def\pb{{\bar{\partial}}}




\def\s{\lies}


\def\demi{{1\over 2}}


\def\f{\phi}    \def\fb{{\bar \f}}
    
\def\F{\Phi}

\def\a{\alpha}
\def\b{\beta}
  
\def\d{\delta}  
\def\m{\mu}
\def\n{\nu}
\def\r{\rho}
\def\l{\lambda} 

\def\e{\epsilon}

\def\|{\Big|}
\def\({\Big(}   \def\){\Big)}
\def\[{\Big[}   \def\]{\Big]}



\def\paper#1#2#3#4{#1, {\sl #2}, #3 {\tt #4}}

\def\hh{hep-th/}


\def\PLB#1#2#3{Phys. Lett.~{\bf B#1} (#2) #3}
\def\NPB#1#2#3{Nucl. Phys.~{\bf B#1} (#2) #3}
\def\PRL#1#2#3{Phys. Rev. Lett.~{\bf #1} (#2) #3}
\def\CMP#1#2#3{Comm. Math. Phys.~{\bf #1} (#2) #3}
\def\PRD#1#2#3{Phys. Rev.~{\bf D#1} (#2) #3}
\def\MPL#1#2#3{Mod. Phys. Lett.~{\bf #1} (#2) #3}
\def\IJMP#1#2#3{Int. Jour. Mod. Phys.~{\bf #1} (#2) #3}


\def\unlockat{\catcode`\@=11}
\def\lockat{\catcode`\@=12}

\unlockat


\def\newsec#1{\global\advance\secno by1\message{(\the\secno. #1)}
\global\subsecno=0\global\subsubsecno=0\eqnres@t\noindent {\bf\the\secno. #1}
\writetoca{{\secsym} {#1}}\par\nobreak\medskip\nobreak}
\global\newcount\subsecno \global\subsecno=0
\def\subsec#1{\global\advance\subsecno by1\message{(\secsym\the\subsecno.
#1)}
\ifnum\lastpenalty>9000\else\bigbreak\fi\global\subsubsecno=0
\noindent{\it\secsym\the\subsecno. #1}
\writetoca{\string\quad {\secsym\the\subsecno.} {#1}}
\par\nobreak\medskip\nobreak}
\global\newcount\subsubsecno \global\subsubsecno=0
\def\subsubsec#1{\global\advance\subsubsecno by1
\message{(\secsym\the\subsecno.\the\subsubsecno. #1)}
\ifnum\lastpenalty>9000\else\bigbreak\fi
\noindent\quad{\secsym\the\subsecno.\the\subsubsecno.}{#1}
\writetoca{\string\qquad{\secsym\the\subsecno.\the\subsubsecno.}{#1}}
\par\nobreak\medskip\nobreak}

\def\subsubseclab#1{\DefWarn#1\xdef #1{\noexpand\hyperref{}{subsubsection}%
{\secsym\the\subsecno.\the\subsubsecno}%
{\secsym\the\subsecno.\the\subsubsecno}}%
\writedef{#1\leftbracket#1}\wrlabeL{#1=#1}}
\lockat

\def\dbend{\lower3.5pt\hbox{\manual\char127}}


\def\boxit#1{\vbox{\hrule\hbox{\vrule\kern8pt
\vbox{\hbox{\kern8pt}\hbox{\vbox{#1}}\hbox{\kern8pt}}
\kern8pt\vrule}\hrule}}

\def\mathboxit#1{\vbox{\hrule\hbox{\vrule\kern8pt\vbox{\kern8pt
\hbox{$\displaystyle #1$}\kern8pt}\kern8pt\vrule}\hrule}}


\def\inbar{\,\vrule height1.5ex width.4pt depth0pt}

\font\cmss=cmss10 \font\cmsss=cmss10 at 7pt


\lref\simons{ J. Cheeger and J. Simons, {\it Differential Characters and
Geometric Invariants},  Stony Brook Preprint, (1973), unpublished.}

\lref\cargese{ L.~Baulieu, {\it Algebraic quantization of gauge theories},
Perspectives in fields and particles, Plenum Press, eds. Basdevant-Levy,
Cargese Lectures 1983}

\lref\antifields{ L. Baulieu, M. Bellon, S. Ouvry, C.Wallet, Phys.Letters
B252 (1990) 387; M.  Bocchichio, Phys. Lett. B187 (1987) 322;  Phys. Lett. B
192 (1987) 31; R.  Thorn    Nucl. Phys.   B257 (1987) 61. }

\lref\thompson{ George Thompson,  Annals Phys. 205 (1991) 130; J.M.F.
Labastida, M. Pernici, Phys. Lett. 212B  (1988) 56; D. Birmingham, M.Blau,
M. Rakowski and G.Thompson, Phys. Rept. 209 (1991) 129.}

\lref\tonin{ Tonin}

\lref\wittensix{ E.  Witten, {\it New  Gauge  Theories In Six Dimensions},
\hh{9710065}. }

\lref\orlando{ O. Alvarez, L. A. Ferreira and J. Sanchez Guillen, {\it  A New
Approach to Integrable Theories in any Dimension}, hep-th/9710147.}

\lref\wittentopo{ E.  Witten,  {\it  Topological Quantum Field Theory},
\hh9403195, Commun.  Math. Phys.  {117} (1988)353.  }

\lref\wittentwist{ E.  Witten, {\it Supersymmetric Yang--Mills theory on a
four-manifold}, J.  Math.  Phys.  {35} (1994) 5101.}

\lref\west{ L.~Baulieu, P.~West, {\it Six Dimensional TQFTs and  Self-dual
Two-Forms,} Phys.Lett. B {\bf 436 } (1998) 97, /hep-th/9805200}

\lref\bv{ I.A. Batalin and V.A. Vilkowisky,    Phys. Rev.   D28  (1983)
2567\semi M. Henneaux,  Phys. Rep.  126   (1985) 1\semi M. Henneaux and C.
Teitelboim, {\it Quantization of Gauge Systems}
  Princeton University Press,  Princeton (1992).}

\lref\kyoto{ L. Baulieu, E. Bergschoeff and E. Sezgin, Nucl. Phys.
B307(1988)348\semi L. Baulieu,   {\it Field Antifield Duality, p-Form Gauge
Fields
   and Topological Quantum Field Theories}, hep-th/9512026,
   Nucl. Phys. B478 (1996) 431.  }

\lref\sourlas{ G. Parisi and N. Sourlas, {\it Random Magnetic Fields,
Supersymmetry and Negative Dimensions}, Phys. Rev. Lett.  43 (1979) 744;
Nucl.  Phys.  B206 (1982) 321.  }

\lref\SalamSezgin{ A.  Salam  and  E.  Sezgin, {\it Supergravities in
diverse dimensions}, vol.  1, p. 119\semi P.  Howe, G.  Sierra and P.
Townsend, Nucl Phys B221 (1983) 331.}

\lref\nekrasov{ A. Losev, G. Moore, N. Nekrasov, S. Shatashvili, {\it
Four-Dimensional Avatars of Two-Dimensional RCFT},  hep-th/9509151, Nucl.
Phys.  Proc.  Suppl.   46 (1996) 130\semi L.  Baulieu, A.  Losev,
N.~Nekrasov  {\it Chern-Simons and Twisted Supersymmetry in Higher
Dimensions},  hep-th/9707174, to appear in Nucl.  Phys.  B.  }

\lref\WitDonagi{R.~ Donagi, E.~ Witten, ``Supersymmetric Yang--Mills Theory
and Integrable Systems'', hep-th/9510101, Nucl. Phys.{\bf B}460 (1996)
299-334}
\lref\Witfeb{E.~ Witten, ``Supersymmetric Yang--Mills Theory On A
Four-Manifold,''  hep-th/9403195; J. Math. Phys. {\bf 35} (1994) 5101.}
\lref\Witgrav{E.~ Witten, ``Topological Gravity'', Phys.Lett.206B:601, 1988}
\lref\witaffl{I. ~ Affleck, J.A.~ Harvey and E.~ Witten,
        ``Instantons and (Super)Symmetry Breaking
        in $2+1$ Dimensions'', Nucl. Phys. {\bf B}206 (1982) 413}
\lref\wittabl{E.~ Witten,  ``On $S$-Duality in Abelian Gauge Theory,''
hep-th/9505186; Selecta Mathematica {\bf 1} (1995) 383}
\lref\wittgr{E.~ Witten, ``The Verlinde Algebra And The Cohomology Of The
Grassmannian'',  hep-th/9312104}
\lref\wittenwzw{E. Witten, ``Non abelian bosonization in two dimensions,''
Commun. Math. Phys. {\bf 92} (1984)455 }
\lref\witgrsm{E. Witten, ``Quantum field theory, grassmannians and algebraic
curves,'' Commun.Math.Phys.113:529,1988}
\lref\wittjones{E. Witten, ``Quantum field theory and the Jones
polynomial,'' Commun.  Math. Phys., 121 (1989) 351. }
\lref\witttft{E.~ Witten, ``Topological Quantum Field Theory", Commun. Math.
Phys. {\bf 117} (1988) 353.}
\lref\wittmon{E.~ Witten, ``Monopoles and Four-Manifolds'', hep-th/9411102}
\lref\Witdgt{ E.~ Witten, ``On Quantum gauge theories in two dimensions,''
Commun. Math. Phys. {\bf  141}  (1991) 153}
\lref\witrevis{E.~ Witten,
 ``Two-dimensional gauge theories revisited'', hep-th/9204083; J. Geom.
Phys. 9 (1992) 303-368}
\lref\Witgenus{E.~ Witten, ``Elliptic Genera and Quantum Field Theory'',
Comm. Math. Phys. 109(1987) 525. }
\lref\OldZT{E. Witten, ``New Issues in Manifolds of SU(3) Holonomy,'' {\it
Nucl. Phys.} {\bf B268} (1986) 79 \semi J. Distler and B. Greene, ``Aspects
of (2,0) String Compactifications,'' {\it Nucl. Phys.} {\bf B304} (1988) 1
\semi B. Greene, ``Superconformal Compactifications in Weighted Projective
Space,'' {\it Comm. Math. Phys.} {\bf 130} (1990) 335.}
\lref\bagger{E.~ Witten and J. Bagger, Phys. Lett. {\bf 115B}(1982) 202}
\lref\witcurrent{E.~ Witten,``Global Aspects of Current Algebra'',
Nucl.Phys.B223 (1983) 422\semi ``Current Algebra, Baryons and Quark
Confinement'', Nucl.Phys. B223 (1993) 433}
\lref\Wittreiman{S.B. Treiman, E. Witten, R. Jackiw, B. Zumino, ``Current
Algebra and Anomalies'', Singapore, Singapore: World Scientific ( 1985) }
\lref\Witgravanom{L. Alvarez-Gaume, E.~ Witten, ``Gravitational Anomalies'',
Nucl.Phys.B234:269,1984. }

\lref\nicolai{\paper {H.~Nicolai}{New Linear Systems for 2D Poincar\'e
Supergravities}{\NPB{414}{1994}{299},}{\hh 9309052}.}



\lref\baex{\paper {L.~Baulieu, B.~Grossman}{Monopoles and Topological Field
Theory}{\PLB{214}{1988}{223}.}{}\paper {L.~Baulieu}{Chern-Simons
Three-Dimensional and
Yang--Mills-Higgs Two-Dimensional Systems as Four-Dimensional Topological
Quantum Field Theories}{\PLB{232}{1989}{473}.}{}}

\lref\bg{\paper {L.~Baulieu, B.~Grossman}{Monopoles and Topological Field
Theory}{\PLB{214}{1988}{223}.}{}}

\lref\seibergsix{\paper {N.~Seiberg}{Non-trivial Fixed Points of The
Renormalization Group in Six
 Dimensions}{\PLB{390}{1997}{169}}{\hh 9609161}\semi
\paper {O.J.~Ganor, D.R.~Morrison, N.~Seiberg}{
  Branes, Calabi-Yau Spaces, and Toroidal Compactification of the N=1
  Six-Dimensional $E_8$ Theory}{\NPB{487}{1997}{93}}{\hh 9610251}\semi
\paper {O.~Aharony, M.~Berkooz, N.~Seiberg}{Light-Cone
  Description of (2,0) Superconformal Theories in Six
  Dimensions}{Adv. Theor. Math. Phys. {\bf 2} (1998) 119}{\hh 9712117.}}

\lref\cs{\paper {L.~Baulieu}{Chern-Simons Three-Dimensional and
Yang--Mills-Higgs Two-Dimensional Systems as Four-Dimensional Topological
Quantum Field Theories}{\PLB{232}{1989}{473}.}{}}

\lref\beltrami{\paper {L.~Baulieu, M.~Bellon}{Beltrami Parametrization and
String Theory}{\PLB{196}{1987}{142}}{}\semi
\paper {L.~Baulieu, M.~Bellon, R.~Grimm}{Beltrami Parametrization For
Superstrings}{\PLB{198}{1987}{343}}{}\semi
\paper {R.~Grimm}{Left-Right Decomposition of Two-Dimensional Superspace
Geometry and Its BRS Structure}{Annals Phys. {\bf 200} (1990) 49.}{}}

\lref\bbg{\paper {L.~Baulieu, M.~Bellon, R.~Grimm}{Left-Right Asymmetric
Conformal Anomalies}{\PLB{228}{1989}{325}.}{}}

\lref\bonora{\paper {G.~Bonelli, L.~Bonora, F.~Nesti}{String Interactions
from Matrix String Theory}{\NPB{538}{1999}{100},}{\hh 9807232}\semi
\paper {G.~Bonelli, L.~Bonora, F.~Nesti, A.~Tomasiello}{Matrix String Theory
and its Moduli Space}{}{\hh 9901093.}}

\lref\corrigan{\paper {E.~Corrigan, C.~Devchand, D.B.~Fairlie,
J.~Nuyts}{First Order Equations for Gauge Fields in Spaces of Dimension
Greater Than Four}{\NPB{214}{452}{1983}.}{}}

\lref\acha{\paper {B.S.~Acharya, M.~O'Loughlin, B.~Spence}{Higher
Dimensional Analogues of Donaldson-Witten Theory}{\NPB{503}{1997}{657},}{\hh
9705138}\semi
\paper {B.S.~Acharya, J.M.~Figueroa-O'Farrill, M.~O'Loughlin,
B.~Spence}{Euclidean
  D-branes and Higher-Dimensional Gauge   Theory}{\NPB{514}{1998}{583},}{\hh
  9707118.}}

\lref\Witr{\paper{E.~Witten}{Introduction to Cohomological Field   Theories}
{Lectures at Workshop on Topological Methods in Physics (Trieste, Italy, Jun
11-25, 1990), \IJMP{A6}{1991}{2775}.}{}}

\lref\ohta{\paper {L.~Baulieu, N.~Ohta}{Worldsheets with Extended
Supersymmetry} {\PLB{391}{1997}{295},}{\hh 9609207}.}

\lref\gravity{\paper {L.~Baulieu}{Transmutation of Pure 2-D Supergravity
Into Topological 2-D Gravity and Other Conformal Theories}
{\PLB{288}{1992}{59},}{\hh 9206019.}}

\lref\wgravity{\paper {L.~Baulieu, M.~Bellon, R.~Grimm}{Some Remarks on  the
Gauging of the Virasoro and   $w_{1+\infty}$
Algebras}{\PLB{260}{1991}{63}.}{}}

\lref\fourd{\paper {E.~Witten}{Topological Quantum Field
Theory}{\CMP{117}{1988}{353}}{}\semi
\paper {L.~Baulieu, I.M.~Singer}{Topological Yang--Mills Symmetry}{Nucl.
Phys. Proc. Suppl. {\bf 15B} (1988) 12.}{}}

\lref\topo{\paper {L.~Baulieu}{On the Symmetries of Topological Quantum Field
Theories}{\IJMP{A10}{1995}{4483},}{\hh 9504015}\semi
\paper {R.~Dijkgraaf, G.~Moore}{Balanced Topological Field
Theories}{\CMP{185}{1997}{411},}{\hh 9608169.}}

\lref\wwgravity{\paper {I.~Bakas} {The Large $N$ Limit   of Extended
Conformal Symmetries}{\PLB{228}{1989}{57}.}{}}

\lref\wwwgravity{\paper {C.M.~Hull}{Lectures on $\CW$-Gravity,
$\CW$-Geometry and
$\CW$-Strings}{}{\hh 9302110}, and~references therein.}

\lref\wvgravity{\paper {A.~Bilal, V.~Fock, I.~Kogan}{On the origin of
$W$-algebras}{\NPB{359}{1991}{635}.}{}}

\lref\surprises{\paper {E.~Witten} {Surprises with Topological Field
Theories} {Lectures given at ``Strings 90'', Texas A\&M, 1990,}{Preprint
IASSNS-HEP-90/37.}}

\lref\stringsone{\paper {L.~Baulieu, M.B.~Green, E.~Rabinovici}{A Unifying
Topological Action for Heterotic and  Type II Superstring  Theories}
{\PLB{386}{1996}{91},}{\hh 9606080.}}

\lref\stringsN{\paper {L.~Baulieu, M.B.~Green, E.~Rabinovici}{Superstrings
from   Theories with $N>1$ World Sheet Supersymmetry}
{\NPB{498}{1997}{119},}{\hh 9611136.}}

\lref\bks{\paper {L.~Baulieu, H.~Kanno, I.~Singer}{Special Quantum Field
Theories in Eight and Other Dimensions}{\CMP{194}{1998}{149},}{\hh
9704167}\semi
\paper {L.~Baulieu, H.~Kanno, I.~Singer}{Cohomological Yang--Mills Theory
  in Eight Dimensions}{ Talk given at APCTP Winter School on Dualities in
String Theory (Sokcho, Korea, February 24-28, 1997),} {\hh 9705127.}}

\lref\witdyn{\paper {P.~Townsend}{The eleven dimensional supermembrane
revisited}{\PLB{350}{1995}{184},}{\hh9501068}\semi
\paper{E.~Witten}{String Theory Dynamics in Various Dimensions}
{\NPB{443}{1995}{85},}{\hh 9503124}.}

\lref\bfss{\paper {T.~Banks, W.Fischler, S.H.~Shenker,
L.~Susskind}{$M$-Theory as a Matrix Model~:
A~Conjecture}{\PRD{55}{1997}{5112},} {\hh9610043.}}

\lref\seiberg{\paper {N.~Seiberg}{Why is the Matrix Model
Correct?}{\PRL{79}{1997}{3577},} {\hh 9710009.}}

\lref\sen{\paper {A.~Sen}{$D0$ Branes on $T^n$ and Matrix Theory}{Adv.
Theor. Math. Phys.~{\bf 2} (1998) 51,} {\hh 9709220.}}

\lref\laroche{\paper {L.~Baulieu, C.~Laroche} {On Generalized Self-Duality
Equations Towards Supersymmetric   Quantum Field Theories Of
Forms}{\MPL{A13}{1998}{1115},}{\hh  9801014.}}

\lref\bsv{\paper {M.~Bershadsky, V.~Sadov, C.~Vafa} {$D$-Branes and
Topological Field Theories}{\NPB{463} {1996}{420},}{\hh9511222.}}

\lref\vafapuzz{\paper {C.~Vafa}{Puzzles at Large N}{}{\hh 9804172.}}

\lref\dvv{\paper {R.~Dijkgraaf, E.~Verlinde, H.~Verlinde} {Matrix String
Theory}{\NPB{500}{1997}{43},} {\hh9703030.}}

\lref\wynter{\paper {T.~Wynter}{Gauge Fields and Interactions in Matrix
String Theory}{\PLB{415}{1997}{349},}{\hh9709029.}}

\lref\kvh{\paper {I.~Kostov, P.~Vanhove}{Matrix String Partition
Functions}{}{\hh9809130.}}

\lref\ikkt{\paper {N.~Ishibashi, H.~Kawai, Y.~Kitazawa, A.~Tsuchiya} {A
Large $N$ Reduced Model as Superstring}{\NPB{498} {1997}{467},}{\hh
9612115.}}

\lref\ss{\paper {S.~Sethi, M.~Stern} {$D$-Brane Bound States
Redux}{\CMP{194}{1998} {675},}{\hh 9705046.}}

\lref\mns{\paper {G.~Moore, N.~Nekrasov, S.~Shatashvili} {$D$-particle Bound
States and Generalized Instantons}{} {\hh 9803265.}}

\lref\bsh{\paper {L.~Baulieu, S.~Shatashvili} {Duality from Topological
Symmetry}{} {\hh 9811198.}}

\lref\pawu{ {G.~Parisi, Y.S.~Wu} {}{ Sci. Sinica  {\bf 24} {(1981)} {484}.}}

\lref\coulomb{ {L.~Baulieu, D.~Zwanziger, }   {\it Renormalizable
Non-Covariant Gauges and Coulomb Gauge Limit}, {Nucl.Phys. B {\bf 548 }
(1999) 527.} {\hh 9807024}.}

\lref\rcoulomb{ {D.~Zwanziger, }   {\it Renormalization in the Coulomb
gauge and order parameter for confinement in QCD}, {Nucl.Phys. B {\bf 518
} (1998) 237-272.} {}}

\lref\kugoojima{ {T.~Kugo and I. Ojima, }   {\it Local covariant
operator formalism of non-Abelian gauge theories and quark confinement
problem}, {Suppl. Prog. Theor. Phys. {\bf 66 } (1979) 1-130.} {}}

\lref\horne{ {J.H.~Horne, }   {\it
Superspace versions of Topological Theories}, {Nucl.Phys. B {\bf 318
} (1989) 22.} {}}

\lref\sto{ {S.~Ouvry, R.~Stora, P.~Van~Baal }   {\it
}, {Phys. Lett. B {\bf 220
} (1989) 159;} {}{ R.~Stora, {\it Exercises in   Equivariant Cohomology},
In Quabtum Fields and Quantum Space Time, Edited
by 't Hooft et al., Plenum Press, New York, 1997}            }

\lref\dzvan{ {D.~Zwanziger, }   {\it Vanishing of zero-momentum lattice
gluon propagator and color confinement}, {Nucl.Phys. B {\bf 364 }
(1991) 127.} }

\lref\dan{ {D.~Zwanziger},  {\it Covariant Quantization of Gauge
Fields without Gribov Ambiguity}, {Nucl. Phys. B {\bf   192}, (1981)
{259}.}{}}

\lref\danzinn{  {J.~Zinn-Justin, D.~Zwanziger, } {}{Nucl. Phys. B  {\bf
295} (1988) {297}.}{}}

\lref\danlau{ {L.~Baulieu, D.~Zwanziger, } {\it Equivalence of Stochastic
Quantization and the-Popov Ansatz,
  }{Nucl. Phys. B  {\bf 193 } (1981) {163}.}{}}

\lref\munoz{ { A.~Munoz Sudupe, R. F. Alvarez-Estrada, } {}
Phys. Lett. {\bf 164} (1985) 102; {} {\bf 166B} (1986) 186. }

\lref\okano{ { K.~Okano, } {}
Nucl. Phys. {\bf B289} (1987) 109; {} Prog. Theor. Phys.
suppl. {\bf 111} (1993) 203. }

\lref\baugros{ {L.~Baulieu, B.~Grossman, } {\it A topological Interpretation
of  Stochastic Quantization} {Phys. Lett. B {\bf  212} {(1988)} {351}.}}

\lref\bautop{ {L.~Baulieu}{ \it Stochastic and Topological Field Theories},
{Phys. Lett. B {\bf   232} (1989) {479}}{}; {}{ \it Topological Field Theories
And Gauge Invariance in Stochastic Quantization}, {Int. Jour. Mod.  Phys. A
{\bf  6} (1991) {2793}.}{}}

\lref\bautopr{  {L.~Baulieu, B.~Grossman, } {\it A topological Interpretation
of  Stochastic Quantization} {Phys. Lett. B {\bf  212} {(1988)} {351}};
 {L.~Baulieu}{ \it Stochastic and Topological Field Theories},
{Phys. Lett. B {\bf   232} (1989) {479}}{}; {}{ \it Topological Field Theories
And Gauge Invariance in Stochastic Quantization}, {Int. Jour. Mod.  Phys. A
{\bf  6} (1991) {2793}.}{}}

\lref\bautoprr{  {L.~Baulieu, B.~Grossman, } { } {Phys. Lett. B {\bf  212}
{(1988)} {351}};
 {L.~Baulieu}{ },
{Phys. Lett. B {\bf   232} (1989) {479}}{}; {}{  }, {Int. Jour. Mod.
Phys. A {\bf  6} (1991) {2793}.}{}}
\lref\samson{ {L.~Baulieu, S.L.~Shatashvili, { \it Duality from Topological
Symmetry}, {JHEP {\bf 9903} (1999) 011, hep-th/9811198.}}}{}

\lref\halpern{ {H.S.~Chan, M.B.~Halpern}{}, {Phys. Rev. D {\bf   33} (1986)
{540}.}}

\lref\yue{ {Yue-Yu}, {Phys. Rev. D {\bf   33} (1989) {540}.}}

\lref\neuberger{ {H.~Neuberger,} {\it Non-perturbative gauge Invariance},
{ Phys. Lett. B {\bf 175} (1986) {69}.}{}}

\lref\gribov{  {V.N.~Gribov,} {}{Nucl. Phys. B {\bf 139} (1978) {1}.}{}}

\lref\huffel{ {P.H.~Daamgard, H. Huffel},  {}{Phys. Rep. {\bf 152} (1987)
{227}.}{}}

\lref\creutz{ {M.~Creutz},  {\it Quarks, Gluons and  Lattices, }  Cambridge
University Press 1983, pp 101-107.}

\lref\zinn{ {J. ~Zinn-Justin, }  {Nucl. Phys. B {\bf  275} (1986) {135}.}}

\lref\gozzi{ {E. ~Gozzi,} {\it Functional Integral approach to Parisi--Wu
Quantization: Scalar Theory,} { Phys. Rev. {\bf D28} (1983) {1922}.}}

\lref\singer{
 I.M. Singer, { Comm. of Math. Phys. {\bf 60} (1978) 7.}}

\lref\neu{ {H.~Neuberger,} {Phys. Lett. B {\bf 295}
(1987) {337}.}{}}

\lref\testa{ {M.~Testa,} {}{Phys. Lett. B {\bf 429}
(1998) {349}.}{}}

\lref\Martin{ L.~Baulieu and M. Schaden, {\it Gauge Group TQFT and Improved
Perturbative Yang--Mills Theory}, {  Int. Jour. Mod.  Phys. A {\bf  13}
(1998) 985},   hep-th/9601039.}

\lref\ostseil { K.~Osterwalder and E.~Seiler, {\it Gauge field theories on
the lattice} {  Ann. Phys. {\bf 110} (1978) 440}.}

\lref\fradshen { E.~Fradken and S.~Shenker, {\it Phase diagrams of lattice
guage theories with Higgs fields} {  Phys. Rev. {\bf D19} (1979) 3682}.}

\lref\banksrab{ T.~Banks and E.~Rabinovici, {} {  Nucl. Phys. {\bf B160}
(1979) 349}.}

\lref\nielsen{N. K. Nielsen, {\it On The Gauge Dependence Of Spontaneous
Symmetry Breaking In Gauge Theories},
Nucl.\ Phys.\ {\bf B101}, 173 (1975)}

\lref\nadkarni{ S. Nadkarni, {\it The SU(2) adjoint Higgs model in three
dimensions } {  Nucl. Phys. {\bf B334} (1990) 559}.}

\lref\stackteper{ A.~Hart, O.~Philipsen, J.~D.~Stack, and M.~Teper,
{\it On the phase diagram of the SU(2) adjoint Higgs model in 2+1
dimensions } { hep-lat/9612021}.}

\lref\kajantie{K.~Kajantie, M.~Laine, K.~Rummujkainen, M.~Shaposhnikov
{\it 3D SU(N)+adjoint Higgs theory and finite temperature QCD }
{hep-ph/9704416}.}

\lref\batrouni{G. G. Batrouni, G. R. Katz, A. S. Kronfeld, G. P. Lepage,
B. Svetitsky and K. G. Wilson, {} {Phys. Rev. {\bf D32} (1985) 2736}}

\lref\davies{C. T. H. Davies, G. G. Batrouni, G. R. Katz, A. S. Kronfeld,
G. P. Lepage, K. G. Wilson, P. Rossi and B. Svetitsky, {} {Phys. Rev. {\bf
D41} (1990) 1953}}

\lref\fukugita{M. Fukugita, Y. Oyanagi and A. Ukawa, {}
{Phys. Rev. Lett. {\bf 57} (1986) 953;
Phys. Rev. {\bf D36} (1987) 824}}

\lref\kronfeld{A. S. Kronfeld, {\it Dynamics of Langevin Simulation} {Prog.
Theor. Phys. Suppl. {\bf 111} (1993) 293}}

\lref\polyakov{ A.~Polyakov, {} {Phys. Letts. {\bf B59} (1975) 82;
Nucl. Phys. {\bf B120} (1977) 429}; {\it Gauge fields and strings,}
ch. 4 (Harwood Academic Publishers, 1987).}

\lref\thooft{ G.~'t Hooft, {} {Nucl. Phys. {\bf B79} (1974) 276}; {}
Nucl. Phys. {\bf B190} (1981) 455 {}.}

\lref\elitzur{S.~Elitzur, {} {Phys. Rev. {\bf D12} (1975) 3978}}


\lref\baugros{ {L.~Baulieu, B.~Grossman, } {\it A topological Interpretation
of  Stochastic Quantization} {Phys. Lett. B {\bf  212} {(1988)} {351}.}}

\lref\bautop{ {L.~Baulieu}{ \it Stochastic and Topological Field Theories},
{Phys. Lett. B {\bf   232} (1989) {479}}{}; {}{ \it Topological Field Theories
And Gauge Invariance in Stochastic Quantization}, {Int. Jour. Mod.  Phys. A
{\bf  6} (1991) {2793}.}{}}

\lref\bautopr{  {L.~Baulieu, B.~Grossman, } {\it A topological Interpretation
of  Stochastic Quantization} {Phys. Lett. B {\bf  212} {(1988)} {351}};
 {L.~Baulieu}{ \it Stochastic and Topological Field Theories},
{Phys. Lett. B {\bf   232} (1989) {479}}{}; {}{ \it Topological Field Theories
And Gauge Invariance in Stochastic Quantization}, {Int. Jour. Mod.  Phys. A
{\bf  6} (1991) {2793}.}{}}

\lref\samson{ {L.~Baulieu, S.L.~Shatashvili, { \it Duality from Topological
Symmetry}, {JHEP {\bf 9903} (1999) 011, hep-th/9811198.}}}{}

\lref\halpern{ {H.S.~Chan, M.B.~Halpern}{}, {Phys. Rev. D {\bf   33} (1986)
{540}.}}

\lref\yue{ {Yue-Yu}, {Phys. Rev. D {\bf   33} (1989) {540}.}}

\lref\neuberger{ {H.~Neuberger,} {\it Non-perturbative gauge Invariance},
{ Phys. Lett. B {\bf 175} (1986) {69}.}{}}

\lref\huffel{ {P.H.~Daamgard, H. Huffel},  {}{Phys. Rep. {\bf 152} (1987)
{227}.}{}}

\lref\creutz{ {M.~Creutz},  {\it Quarks, Gluons and  Lattices, }  Cambridge
University Press 1983, pp 101-107.}

\lref\zinn{ {J. ~Zinn-Justin, }  {Nucl. Phys. B {\bf  275} (1986) {135}.}}

\lref\shamir{  {Y.~Shamir,  } {\it Lattice Chiral Fermions
  }{ Nucl.  Phys.  Proc.  Suppl.  {\bf } 47 (1996) 212,  hep-lat/9509023;
V.~Furman, Y.~Shamir, Nucl.Phys. B {\bf 439 } (1995), hep-lat/9405004.}}

 \lref\kaplan{ {D.B.~Kaplan, }  {\it A Method for Simulating Chiral
Fermions on the Lattice,} Phys. Lett. B {\bf 288} (1992) 342; {\it Chiral
Fermions on the Lattice,}  {  Nucl. Phys. B, Proc. Suppl.  {\bf 30} (1993)
597.}}

\lref\neubergerr{ {H.~Neuberger, } {\it Chirality on the Lattice},
hep-lat/9808036.}

\lref\neubergers{ {Rajamani Narayanan, Herbert Neuberger,} {\it INFINITELY MANY
    REGULATOR FIELDS FOR CHIRAL FERMIONS.}
    Phys.Lett.B302:62-69,1993.
    [HEP-LAT 9212019]}

\lref\neubergert{ {Rajamani Narayanan, Herbert Neuberger,}{\it CHIRAL FERMIONS
    ON THE LATTICE.}
    Phys.Rev.Lett.71:3251-3254,1993.
    [HEP-LAT 9308011]}

\lref\neubergeru{ {Rajamani Narayanan, Herbert Neuberger,}{\it A CONSTRUCTION
OF LATTICE CHIRAL GAUGE THEORIES.}
    Nucl.Phys.B443:305-385,1995.
    [HEP-TH 9411108]}

\lref\neubergerv{ {Herbert Neuberger,}{\it EXACTLY MASSLESS QUARKS ON THE
    LATTICE.}
    Phys. Lett. B417 (1998) 141-144.
    [HEP-LAT 9707022]}

%

\lref\neubergerw{ {Herbert Neuberger,}{\it CHIRAL FERMIONS ON THE
LATTICE.}
    Nucl. Phys. B, Proc. Suppl. 83-84 (2000) 67-76.
    [HEP-LAT 9909042]}

\lref\zbgr {L.~Baulieu and D. Zwanziger, {\it QCD$_4$ From a
Five-Dimensional Point of View},    Nucl. Phys. {\bf B581} 2000, 604;
hep-th/9909006.}
\lref\bgz {P. A. Grassi, L.~Baulieu and D. Zwanziger, {\it Gauge and
Topological Symmetries in the Bulk Quantization of Gauge Theories},
hep-th/0006036.}
\lref\bulkq {L.~Baulieu and D. Zwanziger, {\it From stochastic
quantization to bulk quantization; Schwinger-Dyson equations and
the S-matrix},    hep-th/0012103.}
\lref\equivstoch {L.~Baulieu and D. Zwanziger, {\it Equivalence of
stochastic quantization and the Faddeev-Popov Ansatz},
Nucl. Phys. B193 (1981) 163-172.}

 \lref\zbsd {L.~Baulieu and D. Zwanziger, {
\it From stochastic quantization to bulk quantization: Schwinger-Dyson
equations and S-matrix QCD$_4$}, hep-th/0012103.}
\lref\cuzwns {A.~Cucchieri and D.~Zwanziger, {\it Numerical study of
gluon propagator and confinement scenario in minimal Coulomb gauge},
hep-lat/0008026.}
\lref\fitgrib {A.~Cucchieri and D.~Zwanziger, {\it Fit to gluon
propagator and Gribov formula},    hep-th/0012024.}
\lref\vanish {D.~Zwanziger, {\it Vanishing of zero-momentum lattice
gluon propagator and color confinement},   Nucl. Phys. {\bf B364}
(1991) 127-161.}
\lref\gribov {V.~N.~Gribov, {\it Quantization of non-Abelian gauge
theories},   Nucl. Phys. {\bf B139} (1978)~1-19.}
\lref\singer {I.~Singer, {\it }   Comm. Math. Phys. {\bf 60}
(1978)~7.}
\lref\feynman {R.~P.~Feynman, {\it The qualitative behavior of
Yang-Mills theory in 2+1 dimensions},   Nucl. Phys. {\bf B188} (1981)
479-512.}
\lref\cutkosky {R.~E.~Cutkosky, {\it}   J. Math. Phys. {\bf 25} (1984)
939; R. E. Cutkosky and K. Wang, Phys Rev. {\bf D37} (1988) 3024; R. E.
Cutkosky, Czech. J. Phys. {\bf 40} (1990) 252.}
\lref\vanbaal{J. Koller and P. van Baal, Nucl. Phys. {\bf B302} (1988)
1;
P. van Baal, Acta Phys. Pol. {\bf B20} (1989) 295;
P. van Baal, Nucl. Phys. {\bf B369} (1992) 259;
P. van Baal and N. D. Hari Dass, Nucl. Phys. {\bf B385} (1992) 185.}




\Title{\vbox
{\baselineskip 10pt
\hbox{hep-th/0107074}
\hbox{LPTHE-00-50}
\hbox{NYU-TH-7.10.01}
 \hbox{   }
}}
{\vbox{\vskip -30 true pt
\centerline{
   }
\medskip
 \centerline{  }
\centerline{Bulk Quantization of Gauge Theories:}
\centerline{Confined and Higgs Phases}
\medskip
\vskip4pt }}
\centerline{{\bf Laurent Baulieu}$^{  \dag     }$
  and  {\bf  Daniel
Zwanziger}$^{ \ddag}$}
\centerline{baulieu@lpthe.jussieu.fr, Daniel.Zwanziger@nyu.edu}
\vskip 0.5cm
\centerline{\it $^{\dag}$LPTHE, Universit{\'e}s P. \& M. Curie (Paris~VI)
et D. Diderot (Paris~VII), Paris,  France,}
\centerline{\it $^{\ddag}$   Physics Department, New York University,
New-York,  NY 10003,  USA}

\medskip
\vskip  1cm
\noindent
We deepen the understanding of the quantization of the Yang--Mills field
by showing that the concept of gauge fixing in 4 dimensions  is replaced in
the 5-dimensional  formulation by a procedure that amounts to an
$A$-dependent gauge transformation.  The 5-dimensional
formulation implements the restriction of the
physical 4-dimensional gluon field to the Gribov region, while being a
local description that is under control of BRST symmetries both of
topological and gauge type.  The ghosts decouple so the Euclidean
probability density is everywhere positive, in contradistinction to the
Faddeev--Popov method for which the determinant changes sign outside the
Gribov region.  We include in our discussion the coupling of the gauge
theory to a Higgs field, including the case of spontaneously symmetry
breaking. We introduce a minimizing functional on the gauge orbit that
could be of interest for numerical gauge fixing in the
simulations of spontaneously broken lattice gauge theories.  Other new
results are displayed, such as the identification of the  Schwinger--Dyson
equation of the five dimensional formulation  in the (singular) Landau
gauge  with that of the ordinary Faddeev--Popov formulation, order by
order in perturbation theory.

\Date{\ }

\def\e{\epsilon}
\def\demi{{1\over 2}}
\def\quart{{1\over 4}}
\def\pa{\partial}
\def\a{\alpha}
\def\b{\beta}
\def\d{\delta}

\def\m{\mu}
\def\n{\nu}
\def\r{\rho}
\def\s{\sigma}
\def\l{\lambda}

\def\y{\eta}
\def\o{\omega}

\def\pb{\bar{\psi}}
\def\F{\Phi}

\newsec{Introduction}

In a previous paper \bulkq, we have described the technique of bulk
quantization, which introduces an additional 5-th time,
which generalizes stochastic time, and shown the
deep relationship between this method and the idea of Topological Field
Theories. The delicate symmetries involved by such theories enforce the
physical picture that physical observables must be confined to a
time-slice of the enlarged space. The BRST symmetry of the theory implies
the formal equivalence of Schwinger-- Dyson equations in the two
formulations.  Reference \bulkq\ gives a direct definition of the physical
$S$-matrix (assuming that it exists) in the 5-dimensional formulation,
together with notion of on-shell particles. Topological invariance ensures
the irrelevance of the details of the evolution along the additional time.
This previous paper stresses the importance of the symmetry of the theory
under reversal of the additional time.  The beauty of the construction is
quite striking. However, in the case of a scalar theory, one hardly finds
advantages for the quantization with an additional time as compared to
the ordinary one. In contrast, for gauge theories, conceptual progress
does occur. The enlargement of the phase space for off-shell processes
solves delicate questions such as the one raised by Gribov a long time ago.
In particular the Euclidean probability density is everywhere positive,
whereas in the 4-dimensional approach the Faddeev-Popov determinant
changes sign outside the Gribov region. Our real interest is thus gauge
theories, which are the subject of this second paper.

	Actually we have in mind the following.  According to the ideas of
Gribov,
solving the question of gauge-fixing in the Yang--Mills theory to reach a
definition of the path integral that is valid non-perturbatively is
equivalent to inventing a method that confines the integration over the
gauge field to a fundamental domain.  Doing this implies that the
gauge-fixing provides non-trivial and essential information about
the gauge field configurations that contribute the functional integral. An
analogous situation holds in string theory.  There one has a free theory,
but the nature of 2D diffeomorphism must be fully accounted for in the
gauge-fixing process, including a consistent analysis of moduli
transformations.  This is known to eventually take into account the full
interaction in the theory, although one has ``merely'' gauge-fixed a free
Lagrangian.  This is an early example where the nature of the interactions
is   determined by the gauge fixing, that is by geometry in the
relevant space.  In the Yang--Mills case, the idea of Gribov was that one
should find a method to restrict the path integral over the gauge field $A$
to one fundamental domain  (with a positive Euclidean weight),
and moreover that this domain could be chosen in
such a way that: (1) $A$ is transverse, and (2) the operator
$-\pa_\mu D_\m(A)$ is positive, that is, all its eigenvalues are positive
for every configuration $A$ in the domain \gribov.  These two conditions
define a (larger)  region known as ``the Gribov region''.  As a
consequence of these conditions, one finds \vanish\ that
the gluon propagator $D(k)$ in the Landau gauge cannot exhibit a pole in
$k$ at $k=0$, and in fact $D(k)$ {\it vanishes} at $k = 0$.  Computer
simulations have recently been shown to sustain this property \cuzwns,
\fitgrib.)  The position of the pole in the transverse part of
the gluon propagator is independent of the gauge parameters, by virtue of
the Nielsen identities \nielsen, so if this pole is absent in the Landau
gauge it is absent in all gauges.  The point of view that we adopt in this
paper is that bulk quantization is a consistent and operational
formulation in 5 dimensions of a quantum field theory  in 4 dimensions
that automatically satisfies the Gribov condition. It follows that there
can be no free massless gluons in the resulting theory.  This is a first
and crucial step toward proving confinement.\foot{In Gribov's original
formulation, a long-range ``force'' that confines all colored particles is
provided in the Coulomb gauge by the long range of the $A_4$--$A_4$
correlator.  In the present 5-dimensional formulation it is provided by the
$A_5$--$A_5$ correlator. The heuristic arguments for
confinement in the Coulomb gauge say that the field $A_4$ carries an
infinite range instantaneous anti-screening force. Of course this argument
is spoiled by the fact that the Coulomb gauge is not well-defined at the
non-perturbative level in the Faddeev-Popov formulation because of the
existence of Gribov copies which cannot be eliminated by a local
action.  This problem is overcome in the 5-dimensional formulation, and
moreover the field $A_5$ has engineering dimension 2. This may eventually
allow a rigorous proof of confinement by using the instantaneous force in
the fifth dimension that is carried by $A_5$ in the Landau gauge
limit.}  Essential properties such as the existence of a mass gap and
bound states remain very difficult, but one may hope that the new
framework of bulk quantization will bring new hints for establishing them.

 Thus, prior to any investigation of its dynamics, the necessity of a
consistent gauge-fixing (in reality, the introduction of a ``drift
force" tangent to gauge orbits) implies that the massless gluon simply
cannot appear in the spectrum -- although it plays an essential role as a
parton -- simply because there is no room non-perturbatively for all
Fourier components of an asymptotic massless field within the Gribov
region.  Further development of Gribov's ideas may be found in
\rcoulomb.  Actually, there have been other attempts to understand
confinement from a  geometrical point of view \kugoojima.

	Similar questions arise when gauge theories are coupled coupled to a
Higgs field.  Perturbatively it seems that the gauge boson can acquire a
mass and become part of the physical spectrum.  However non-perturbatively
there is no clear difference between these two phases because they may be
continuously connected, and the status of the gauge boson as an
elementary particle or bound state is at issue \thooft, \polyakov,
\nadkarni, \stackteper, \kajantie.  We propose a gauge-fixing appropriate
to the Higgs phase in the 5-dimensional formalism which is valid
non-perturbatively.  It selects a direction for the Higgs field in a way
that is consistent with Elitzur's theorem \elitzur.  Moreover it has the
advantage that it may be used in lattice simulations of the Higgs phase
where it may be implemented by a numerical minimization.

	Ideas similar to Gribov's have been developed by Feynman \feynman\ and
Singer \singer.  They have been implemented in concrete dynamical
calculations by Cutkosky and co-workers \cutkosky\ and by van Baal and
co-workers \vanbaal\ in a Hamiltonian formulation of the 4-dimensional
theory, keeping a small number of modes.  A reasonable hadron
spectrum results from the boundary identification of the fundamental
modular region, which confirms the validity of Gribov's approach to
confinement.  However it has proven to be an extremely difficult problem
to carry out this program to a higher degree of accuracy precisely
because, in the 4-dimensional formalism, the boundary of the fundamental
modular region is not provided by the Faddeev-Popov
procedure, and must be found ``by hand", by non-perturbative
calculations.  On the other hand, as explained in \bgz,
and as discussed in more detail below in sec. 3, the 5-dimensional
formulation automatically restricts the
physical 4-dimensional connection to the Gribov region, while being a
local gauge theory that is under control of BRST symmetries both of
topological and gauge type.   This suggests that the Gribov program is
truly realizable  in the context of a local quantum field theory.

This paper is organized as follows.	Sec. 2, which is continued in
Appendix A, contains a pedestrian step-by-step construction of the action
of the 5-dimensional formulation as an alternative to the geometrical
construction given in \bgz. It starts from the formalism that we developed
in the preceding paper devoted to theories that are not of gauge type
\bulkq. It is shown that the concept of {\it gauge fixing}  is replaced
in the 5-dimensional  formulation by a procedure that amounts to an
$A$-dependent {\it gauge transformation}, and one avoids by construction
the objection of Singer~\singer. We show in Appendix B that the Jacobian
of this gauge transformation is an infinite constant, independent of $A$,
which cancels the divergent volume of the gauge group. In Appendix A 
we introduce the BRST-operator $w$ that codifies the 5-dimensional gauge
invariance, with results summarized in sec. 2.  In sec. 3a we show the
equivalence to the previous approach \bgz, and provide a dictionary that
relates the fields introduced here and there.  In sec. 3b we show that
the ghost fields decouple because their field equations are parabolic so
the ghost propagators are retarded and all ghost loops vanish.  In
sec.~4, we indicate how the restoring force along gauge orbits forbids
the existence of massless gluons, independently of the details of the
confining force. Section 5 is devoted to the case of the coupling of the
gauge field to a Higgs field, and we generalize the mechanism that is at
work in the confining phase to the Higgs phase. We have relegated other
results to appendices. In Appendix C, we give a new way of showing the
perturbative equivalence of the 4- and 5-dimensional formulations for
gauge-invariant quantities. It is an alternative to the old proof
\equivstoch. It also explains how the non-pertubatively ill-defined
Faddeev--Popov ghost of the 4D formulation can be extracted (in a
non-local way) from the well-defined topological ghost of the 5D
formulation, in a singular gauge. In Appendix D, we establish the
invariance of a gauge theory under  reversal of  the 5th-time, which
generalizes the case of a theory of non-gauge type.  In Appendix E we
present a semi-classical treatment of the Higgs mechanism in the
5-dimensional formulation.

\newsec{ Step-by-step determination of the TQFT of a gauge theory}

\subsec{ Step 1:  Scalar-field type quantization}

In 4 dimensions, one considers an SU(N) gauge field $A_{\m}^a(x)$, with
Yang-Mills action $S = S_{\rm YM} = \quart \int d^4x F_{\m\n}^2$, where
$F_{\m
\n}^a = \p_\m A_\n^a - \p_\n A_\m^a + f^{abc} A_\m^b A_\n^c$, and  $ { {\d
S} \over {\d A_\m^a} } = - (D_\l F_{\l\m})^a$.  The
gauge-covariant derivative is defined by $(D_\m \omega)^a \equiv \p_\m
\omega^a + f^{abc} A_\m^b \omega^c$.

 To derive the 5-dimensional formulation in its simplest
expression, we start with some results contained in our
paper devoted to the case of theories non-gauge type
\bulkq\ and do the minimal hypothesis that the ungauge-fixed theory
corresponds to the pioneering formulation of Parisi and Wu \pawu,
as developed in \gozzi\ and \zinn. (This first step is formal because of
divergences associated with the infinite volume of the gauge group.)
It prescribes that, corresponding to the 4-dimensional Euclidean field
$A_\mu(x)$, is the quartet
$ A_\mu(x,t),\psi _\mu(x,t), \bar \psi_\mu(x,t), \pi_\mu(x,t) $
on which a topological BRST-operator acts according to
\eqn\sona{\eqalign{
sA_\m & = \psi_\m,  \ \ \ \ \ s\psi_\m = 0   \cr
s\pb_\m & = \pi_\m, \ \ \ \ \ \  s\pi_\m = 0,
}}
and,  moreover, corresponding to the classical Yang-Mills action
$S_{\rm YM}$ is the BRST-exact bulk action
\eqn\gact{\eqalign{
 I_{\rm YM}
& \equiv \int dt \ d^4x \ s[ \pb_\m (\p_t A_\m
+ \d S_{\rm YM} / \d A_\m + \pi_\m)].
}}
We have $\d S_{\rm YM} / \d A_\m = - D_\l F_{\l\m}$,
where $F_{\l\m} = \p_\l A_\m - \p_\m A_l + [A_\l, A_\m]$
is the Yang-Mills field tensor.
The field $\pi_\m$ (which would be written $b_\m$ in the notation of
\bulkq)  is the momentum density canonical to $A_\m$ in the 5-dimensional
theory. Indeed,  upon expansion we obtain
\eqn\xgact{\eqalign{
{I_{\rm YM}} = \int dt \ d^4x \ \Big[ & \ \pi_\m \Big(\p_t A_\m
 - D_\l F_{\l\m} + \pi_\m \Big) \cr
& - \ \pb_\m
\Big(\p_t \psi_\m   - D_\l (D_\l \psi_\m - D_\m\psi _\l)
- [\psi_\l, F_{\l\m}] \Big) \ \Big],  }}
and we have $ \pi_\m = \p {\cal L} / \p \dot{A}_\m$, where
$\dot{A}_\m \equiv \p_t A_\m$.
Similarly we have
$ \pb_\m = \p {\cal L} / \p \dot{\psi}_\m$, where the left-derivative is
taken.  A consequence of this action is that on a given time slice the
formal ungauge-fixed Schwinger-Dyson equations of $S_{\rm YM}$ are
satisfied, as in theory of non-gauge type~\bulkq.

\subsec{Step 2: Normalization of the path integral by gauge
transformation}

	The action $I_{\rm YM}$  inherits from $S_{\rm YM}$
invariance under local $4$-dimensional gauge transformations $g(x)$
under which the fields transform according to
\eqn\gtrans{\eqalign{
 A_\m & \to {^gA}_\m = g^{-1} A_\m g + g^{-1} \p_\m g   \cr
\psi_\m & \to {^g\psi}_\m = g^{-1} \psi_\m g    \cr
\pb_\m & \to {^g\pb}_\m = g^{-1} \pb_\m g    \cr
\pi_\m & \to {^g\pi}_\m = g^{-1} \pi_\m g.
}}
Consequently the action $I_{\rm YM}$ provides no convergence for the
longitudinal modes. We must cure this problem.

	Consider now  a local gauge transformation that is also
$t$-dependent, $g = g(x, t)$.  This is clearly a symmetry
transformation for gauge-invariant observables
${\cal O}({^g}A) = {\cal O}(A)$, in
particular for those that depend on $A_\m(x,t)$ at a fixed time,
$A_\m(x,0)$ say.\foot {For a pure gauge theory without quarks, the
time $t$ has a stochastic interpretation and corresponds to the number
of sweeps in a Monte Carlo calculation.  The gauge transformation
$g(x,t)$ corresponds to making a gauge transformation after each
sweep.  The functional dependence of $g(x,t)$ on $A$ corresponds to
choosing $g(x,t)$ to depend on $A(x,t)$, as is common practice when $g
= g[A]$ is chosen by a minimization process.}  However the action
$I_{\rm YM}$ is not invariant under transformations $g(x,t)$ because
of the terms involving time derivatives.  They transform according to
\eqn\ggtrsfm{\eqalign{
\p_t A_\m & \to \p_t({^gA}_\m) = g^{-1}(\p_tA_\m - D_\m v)g \cr
\p_t \psi_\m  & \to \p_t({^g\psi}_\m)
=g^{-1}( \p_t \psi_\m - [\psi_\m, v])g \ ,
}}
where $[X, Y]^a \equiv f^{abc}X^b Y^c$, and
$v \equiv - \p_tg g^{-1} = g\p_tg^{-1}$ .  Under this gauge transformation
$I_{\rm YM}$ becomes
\eqn\ngact{\eqalign{
{\hat{I}_{\rm YM}} = \int dt \ d^4x \ s[ \ \pb_\m (\p_t A_\m - D_\m v
- D_\l F_{\l\m} + \pi_\m) \ ].  }}
 This action is physically equivalent
to $I_{\rm YM}$.  Moreover given any $v(x,t)$, we may solve for
$g(x, t)$, so $v(x,t)$ is a function at our disposal.  We shall
in fact choose
\eqn\choose{\eqalign{
v = a^{-1}\p_\m A_\m,
}}
as was also made by \dan,\halpern and \bautoprr.  With this choice, that
is actually  compulsory if renormalizability by power counting is required
in the five dimensional quantum field theory, the
action ${\hat{I}_{\rm YM}}$ provides convergence for all modes including
the longitudinal modes, as we shall see.  In doing
so we are merely choosing a gauge {\it transformation},
$v = - \p_tg g^{-1}$, but no gauge {\it fixing} is done so the issue of
Gribov copies does not arise.\foot {This was pointed out many years
ago in the context of the Langevin equation \dan. Here we obtain the
same result by the more conventional functional integral methods.}
Once $v$ is determined, so is its $s$-transform $sv$.  The transformed
action with $v = a^{-1}\p_\m A_\m$ is given after expansion by
\eqn\egact{\eqalign{
{\hat{I}_{\rm YM}} = \int dt \ d^4x \ \Big[ & \ \pi_\m \Big(\p_t A_\m -
a^{-1}D_\m \p_\l A_\l - D_\l F_{\l\m} + \pi_\m \Big) \cr & - \ \pb_\m
\Big(\p_t \psi_\m - a^{-1}D_\m \p_\l \psi_\l - a^{-1}[\psi_\m, \p_\l
A_\l] \cr &\ \ \ \ \ \ \ \ \ \ \ - D_\l (D_\l \psi_\m - D_\m\psi _\l)
- [\psi_\l, F_{\l\m}] \Big) \ \Big].  }}

 To see that this action provides convergence for the longitudinal
modes consider its quadratic part,
\eqn\qegact{\eqalign{
(\hat{I}_{\rm YM})_0
=  \int dt \ d^4x \ \Big[ & \ \pi_\m \Big(\p_t A_\m - a^{-1}\p_\m \p_\l
A_\l - \p_\l (\p_\l A_\m - \p_\m A_\l) + \pi_\m \Big)   \cr
 & - \ \pb_\m \Big(\p_t \psi_\m - a^{-1} \p_\m \p_\l \psi_\l
    - \p_\l (\p_\l \psi_\m - \p_\m\psi _\l)
   \Big)  \ \Big].
}}
From it we obtain the free propagators in momentum space
\eqn\frprop{\eqalign{
& D^{AA, {\rm tr}} = 2[\o^2 + (k^2)^2]^{-1},   \ \ \ \ \ \ \ \ \
D^{AA, {\rm long}} = 2[\o^2 +a^{-2} (k^2)^2]^{-1}  \cr
& D^{\psi \pb, {\rm tr}} = [i\o + k^2]^{-1}, \ \ \ \ \ \ \ \ \ \ \ \ \ \ \
D^{\psi \pb, {\rm long}} = [i\o + a^{-1}k^2]^{-1},
}}
and by $s$-invariance $D_{\l \m}^{A b} = - D_{\l \m}^{\psi \pb}$.  The
parameter $a^{-1}$ provides convergence of the longitudinal modes, as
asserted, and $a = 0$ is the Landau-gauge limit.  We take the
parameter $a^{-1} > 0$.

	 It may seem paradoxical that regularization of the
longitudinal modes, which requires division by the infinite volume of
the gauge orbit, has been be achieved by a gauge transformation, which is
the interpretation that we gave in this section.  The answer to this is
that  gauge transformations $g$ leave the functional measure ${\cal D}A$
invariant only when $g$ is independent of the variable $A$.  However if $g
= g[A]$ is a functional\foot {The transformations \gtrans\ of
$\psi_\m$ and $\pi_\m$ hold only when $g$ is independent of $A$.
Otherwise they are given by ${^g\psi}_\m = s \ {^gA}_\m$ and ${^g\pi}_\m
= s \ {^g\pb}_\m$.}  of $A$, then the Jacobian $J$ of the
transformation $A \to {^gA}$ is not necessarily unity.  In Appendix B,
we will calculate $J$ for the transformation with $v = - \p_tgg^{-1} =
a^{-1}\p_\m A_\m$, and find that it is an infinite constant, independent
of $A$.  This is an essential point, for if the Jacobian of the
regularizing gauge transformation were
$A$-dependent, $J = J[A]$, there would be additional corrections
to the action.  Clearly the mechanism of regularization by gauge
transformation is quite different from ordinary gauge fixing which
requires choosing a point on each gauge orbit, and for this reason the
Gribov problem does not arise.

\subsec{Step 3: Introduction of the 5th component $A_5$}

	If one introduces the notation
\eqn\fdnot{\eqalign{
x_5  \equiv t, \ \ \ \ \ \
A_5  \equiv v, \ \ \ \ \ \
\psi_5  \equiv sA_5,
}}
one recognizes that
$ \p_tA_\m - D_\m v = \p_5A_\m - \p_\m A_5 - [A_\m, A_5]
= F_{5\m}$
is a component of the Yang-Mills field tensor in 5
dimensions. Regarded as a function of the new variables, the action
$\hat{I}_{\rm YM}(A_\m, A_5, \psi_\m, \psi_5, \pb_\m, \pi_\m)$
reads
\eqn\gcgact{\eqalign{
 \hat{I}_{\rm YM}
\equiv  \int d^5x \ s[  \ \pb_\m (F_{5\m}
- D_\l F_{\l\m} + \pi_\m)      \ ].
}}
It is manifestly invariant under the 5-dimensional gauge
transformation $g(x,t)$ that depends both on $x$ and $t$, under which
the fields transform according to \gtrans\ supplemented by
\eqn\sgtrans{\eqalign{
 A_5 & \to {^gA}_5 = g^{-1} A_5 g + g^{-1} \p_5 g   \cr
\psi_5 & \to {^g\psi}_5 = g^{-1} \psi_5 g.
}}

	We have seen that we may regularize the
longitudinal modes by choosing
\eqn\xgfix{\eqalign{
A_5 = a^{-1}\p_\n A_\n,
}}
without encountering the Gribov problem for the physical variables
$A_\m$, with $\m = 1,...4$. The new notation reveals that the
regularization of the 4-dimensional gauge invariance by gauge
transformation resembles a linear gauge-fixing condition of the
5-dimensional gauge symmetry by
$aA_5 = \p_\m A_\m$. This relation implies
\eqn\psifive{\eqalign{
\psi_5 = a^{-1}\p_\n \psi_\n.
}}

	Note that $A_5$ itself appears in the gauge condition
rather than its derivative $\p_5 A_5$.
In this respect the gauge-fixing of the 5-dimensional theory
resembles the axial gauge for which
the ghosts decouple because, as we shall see,
the Faddeev-Popov determinant is
known to be trivial.  However whereas the axial gauge in 4 dimensions
is ambiguous, the 5-dimensional theory is well-defined and renormalizable.
Moreover this gauge condition does not violate Singer's theorem
because t extends over an
infinite interval whereas Singer's theorem applies to compact space-time
\singer.  The gain over the conventional Faddeev-Popov formulation is
enormous because the ghosts decouple, as will be shown below.
As a result, the Euclidean weight, after elimination of auxiliary fields,
is positive everywhere.  By contrast, in the 4-dimensional approach the
Faddeev-Popov determinant changes sign outside the Gribov region.

\def\mb{{\bar m}}

We automate the conditions $aA_5=\p_\mu A_\m$
and $a\psi_5=\p_\mu \psi_\m$
by adding an action
\eqn\gfixa{\eqalign{
\hat{I}_{\rm gf}  \equiv
 \int d^5x \ [ \  l(a A_5 - \p_\n A_\n)
- \mb(a \psi_5 - \p_\n \psi_\n) \  ].
}}
that contains two Lagrange multiplier fields $l$ and $\mb$, one for each of
these the conditions. To maintain $s$-invariance and to
keep $s$ trivial, in the sense that it acts on an
elementary field to produce another elementary field rather than a
composite field, we arrange the new variables and their Lagrange
multipliers into a quartet
$(A_5, \psi_5, \mb, l)$,  like $(A_\m, \psi_\m, \pb_\m, \pi_\m)$, within
which $s$ acts according to
\eqn\sonpb{\eqalign{
&sA_5 = \psi_5, \ \ \ \ \ \ \ s\psi_5 = 0, \cr
&s\mb = l, \ \ \ \ \ \ \ \ \ \ \ sl = 0, }}
as in \sona.  (We use the notation $\mb$ and $l$ -- rather than $\pb_5$ and
$\pi_5$ --
for these Lagrange multiplier fields that enforce time-independent
constraints, to distinguish them from the four $\pb_\m$ and $\pi_\m$,
that impose time-dependent equations of motion.)  The new action may be
written in the $s$-exact form
\eqn\gfixsa{\eqalign{
\hat{I}_{\rm gf}  = \int d^5x \ s[ \  \bar{m}(a A_5 - \p_\n A_\n)
 \  ],
}}
and the total action
\eqn\tgfix{\eqalign{
\hat{I} & \equiv \hat{I}_{\rm YM} + \hat{I}_{\rm gf}   \cr
        & = \int d^5x \ s[
 \ \pb_\m (F_{5\m} - D_\l F_{\l\m} + \pi_\m)
+ \mb (a A_5 - \p_\n A_\n) \ ]
}}
is equivalent to the action \egact.  As it stands, this action
does not provide easy access to the Ward identities that express the
5-dimensional gauge invariance of $\hat{I}_{\rm YM}$.  This will be
done by the introduction of a second BRST operator $w$ that encodes
gauge invariance.

\subsec{Summary: s and w on all fields and 5-dimensional action}

\def\pbi{{\bar \psi }}
	Steps 4 and 5 are somewhat lengthy and are consigned to Appendix A.  We
summarize here the results of steps 4 and 5.  There are two BRST
operators: the $s$-operator, introduced above, that is topological,
and the $w$-operator that encodes gauge invariance.  They are algebraically
consistent in the sense that $s^2 = w^2 = sw + ws = 0$.

	The action of $s$ and of $w$ on all fields is given by
\eqn\sonall{\eqalign{
sA_\n  & = \psi_\n \ \ \ \ \ \ \ s\psi_\n = 0 \
\ \ \ \ \ s\bar{\psi}_\n  = \pi_\n \ \ \ \ \ \
\ \ s\pi_\n = 0    \cr
sA_5  & = \psi_5 \ \ \ \ \ \ \ s\psi_5 = 0 \
\ \ \ \ \ \  s\bar{m}  = l \ \ \ \ \ \ \ \ \ \
\ \ sl = 0    \cr
s\l  & = \m \ \ \ \ \ \ \ \ \ \  s\m = 0 \
\ \ \ \ \ \ \ s\bar{\m}  = \bar{\l} \ \ \ \ \ \ \ \
\ \ s\bar{\l} = 0     \cr
s\o  & = \phi \ \ \ \ \ \ \ \ \ \  s\phi = 0 \
\ \ \ \ \ \ \ s\bar{\phi}  = \bar{\o} \ \ \ \ \ \ \ \
\ \ s\bar{\o} = 0;
}}
\eqn\wonall{\eqalign{
& w A_\n  = D_\n \l \ \ \ \ \ \ \ \ \ \
w \psi_\n = - [\l, \psi_\n] - D_\n \m \ \ \ \
w \pbi_\n = - [\l, \pbi_\n] \ \  \ \ \
w \pi_\n = - [\l, \pi_\n] + [\m, \pbi_\n]    \cr
& w A_5  = D_5 \l \ \ \ \ \ \ \ \ \ \ \
w \psi_5 = - [\l, \psi_5] - D_5 \m \ \ \ \ \
w \bar{m} = 0 \ \  \ \ \  \ \ \ \ \ \ \ \ \ \
w l = 0    \cr
& w \l  = - \demi[\l, \l] \ \ \ \ \ \ \ \
w\m = - [\l, \m]   \ \ \ \ \ \ \ \ \ \ \ \ \ \ \ \
w\bar{\m} = \bar{m} \ \ \ \ \ \ \ \ \ \ \ \ \ \ \
w \bar{\l} = - l      \cr
& w \o = - [\l, \o] -\m    \ \ \ \
w \phi = - [\l, \o] + [\m, \o] \ \ \ \ \ \
w \bar{\phi} = - [\l, \bar{\phi}] \ \ \ \ \ \ \ \
w \bar{\o} = - [\l, \bar{\o}] + [\m, \bar{\phi}].
}}
The algebra of $s$ and $w$ closes on the fields of the first three lines.
The last quartet is not needed for algebraic consistency  but is
needed to construct an action that is both $s$- and
$w$-invariant.

	Associated to the symmetry
generators $s$ and $w$ are independently conserved ghost numbers $N_s$
and $N_w$ which are increased by unity by the action of $s$ and $w$.
We make the following ghost number assignment, indicated by the
superscripts $(N_s, N_w)$, consistent with this and
with \sonall\ and \wonall:
\eqn\diagra{\eqalign{\matrix{
\matrix{
    &  \ \   & \pb_\n^{(-1,0)} & \to & \pi_\n^{(0,0)}
\  & A_\n^{(0,0)} & \to & \psi_\n^{(1,0)}  &     \cr
     &  & \mb^{(-1,0)} & \to & l^{(0,0)}
& A_5^{(0,0)} & \to & \psi_5^{(1,0)}         \cr
 & \nearrow  &   & \nearrow &     \cr
  \bar{\m}^{(-1,-1)} & \to & \bar{\l}^{(0,-1)}  &&&&&
\l^{(0,1)} & \to & \m^{(1,1)}   \cr
\fb^{(-2,0)} & \to  & \bar{\o}^{(-1,0)} & & & & &
\o^{(1,0)} & \to & \f^{(2,0)}.
 } }}}
Each column corresponds to a fixed value of the total ghost
number $N \equiv N_s + N_w$.   Fields with even $N$ are bosonic;
otherwise they are fermionic.  Each row corresponds to a single quartet
within which $s$ acts as indicated by the horizontal arrows $\to$.  The two
northeast arrows $\nearrow$ indicate the action of $w$, but only where it
produces the elementary Lagrange multiplier fields $\bar{m}$ and $-l$.
(Otherwise $w$ produces a composite field.)  This is the minimum number of
fields that is required to construct an action that is both $s$- and
$w$-invariant and that is physically equivalent to the preceding action.

	We also introduce the composite fields,
\eqn\covb{\eqalign{
\pi^\star_\n \equiv \pi_\n + [\o, \pbi_\n]
}}
\eqn\covpsi{\eqalign{
\psi^\star_\m \equiv \psi_\m - D_\m \o, \ \ \ \ \ \
\psi^\star_5 \equiv \psi_5 - D_5 \o,
}}
that are $w$-covariant,
$w \pi^\star_\n = - [\l, \pi^\star_\n]$.
$w \psi^\star_\m = - [\l, \psi^\star_\m]$,
$w\psi^\star_5 = - [\l, \psi^\star_5]$.

	A consistent bulk action for gauge fields is provided by
\eqn\cgact{\eqalign{
  I & \equiv  \int d^5x \ \Big[ \ s\Big( \pbi_\m (F_{5\m} - D_\l
F_{\l\m}  + \pi_\m + [\o, \pb_\m] )
+  \bar{\f}[ \  a' (\psi_5 - D_5 \o)
- D_\m (\psi_\m - D_\m \o)] \Big)    \cr
 & \ \ \ \ \ \ \ \ \ \ \ \
+ sw \Big(  \ \bar{\m}(a A_5 - \p_\n A_\n) \ \Big) \ \Big].
}}
It is $s$-exact and $w$-invariant, $wI = 0$.  All terms except the last
are in the cohomology of $w$, and the last term is $w$-exact. The
expansion of the various terms in this action is given in Appendix A.

\newsec{Equivalence to the previous approach and decoupling of ghost
fields}

\subsec{Equivalence to the geometrical approach}

	Remarkably, the action and fields that have just been derived agree
precisely with the corresponding quantities of \bgz\ which was
obtained by quite different geometrical reasoning. These fields were
displayed in the following pyramidal diagram:
\eqn\diagram{\eqalign{\matrix{
                        \matrix{&&&&A^{(00)}_\m
,A^{(00)}_5
\cr
                       &&&\swarrow&&\ \cr
             &&\Psi_\mu^{(1,0)}
,\Psi_5^{(1,0)}
&&&&\bar{\Psi}_\m^{(-1,0)},
\cr
                    &&  c^{(1,0)},\l^{(0,1)}
&&&&  \bar{c} ^{(-1,0)}\ \bar  \lambda   ^{(0,-1)}
 \cr
                       &\swarrow&&\ &&\swarrow&&\nwarrow \cr
                       \Phi^{(2,0)},
&&&&
b_\mu^{(0,0)}&&&&\bar{\Phi}^{(-2,0)},
\cr
\mu^{(1,1)}
%
%
%
&&&&
l ^{(0,0)}&&&&\  \bar{\mu}^{(-1,-1)}
\cr
&\ &&\  \  &&&&    \swarrow  \cr
\cr
                   &&
&&&&
 \
\bar \eta ^{(-1,0)},     \cr
                         && \ &&&& \cr } }}}
To exhibit the correspondence between  these fields
and the ones in the present article requires a
non-linear field redefinition to provide fields that transform
gauge-covariantly under $w$.  For this purpose we also need ``adjusted"
fields $\f^\star$ and $\bar\o^\star$ that transform gauge-covariantly,
\eqn\gcfi{\eqalign{
  \f^\star & \equiv \f + \demi [\o, \o]
    \ \ \ \ \ \ \ \  \ \ \ \ \
w\f^\star  = - [\l, \f^\star]     \cr
  \bar\o^\star &\equiv \bar{\o} + [\o, \fb]
\ \ \ \ \ \ \ \ \ \ \ \ \ \ \
w\bar\o^\star  = - [\l, \bar\o^\star].
}}
The correspondences are given by
\eqn\corres{\eqalign{
A_\n & = A_\n; \ \ \ \ \ \ \ \ \ \ \ \ \ \ \ \ \
\ \ \ \ \    \Psi_\n  = \psi_\n^\star = \psi_\n - D_\n \o \cr
\bar{\Psi}_\n & = \pb_\n;    \ \ \ \ \ \
\ \ \ \ \ \ \ \ \ \ \ \ \ \ \ \ \
b_\n = \pi_\n^\star = \pi_\n + [\o,\pb_\n]\cr
A_5 & = A_5;  \ \ \ \ \ \ \ \ \ \ \ \ \ \ \ \ \ \ \ \ \ \ \
\Psi_5 = \psi_5^\star = \psi_5 - D_5 \o \cr
\bar c & = \mb;  \ \ \ \ \ \ \ \ \ \ \ \ \ \ \
\ \ \ \ \ \ \ \ \ \     l = l     \cr
\l & = \l; \ \ \ \ \ \ \m = \m; \ \ \
\ \ \ \ \ \ \ \bar\m = \bar\m; \ \ \ \ \ \ \ \ \ \ \bar\l = \bar\l
\cr
c & = \o;  \ \ \ \ \ \ \ \ \ \ \ \ \ \ \ \ \ \ \ \ \ \ \ \ \
\F  = \f^\star  \equiv \f + (1/2)[\o, \o]     \cr
\bar\F & = \bar\f;
\ \ \ \ \ \ \ \ \ \ \ \ \ \ \ \ \ \ \ \ \ \ \ \ \ \
  \bar{\eta}  = \bar\o^\star \equiv \bar{\o} + [\o, \fb]
}}
(where upper and lower cases are distinguished).
When expressed in terms of the new variables, the
action \cgact\ is
the action of \bgz\ where its renormalizability and other properties
are established.  Because the field redefinition
$\Psi_5 = \psi_5^\star = \psi_5 - D_5\o$
involves a time derivative, dynamical and non-dynamical field
equations become interchanged in the action, so the non-dynamical
Lagrange multiplier $\mb$ becomes the dynamical Lagrange multiplier
$\bar c$ and conversely for $\bar\o$ and $\bar\eta$.

	Consistency of the construction is revealed by  the  geometrical
formula:
 \eqn\topgsym{\eqalign{
(d+s+w) (A+c + \l) +{1\over 2}[A+c+\l,A+\o+\l]
=F+\Psi+\Phi         \cr
 (d+s+w) (F+\Psi+\Phi) + [A+c+\l,F+\Psi+\Phi]=0        \cr
(s+w) \bar \Phi +
 [c+\l,\bar\Phi]=\bar \eta          \cr
(s+w) \bar \eta +
 [c+\l,\bar\eta]=[\Phi,\bar\Phi] \cr (s+w) \bar \mu =\bar c+\bar \l
 \cr
(s+w) (\bar c+\bar \l )=0
}}
which implies that one has
 $(s+w)^2=0$ by construction. Equation~\topgsym\ is typical of a
topological gauge symmetry. Fields like $\mu$ and $l$ are
 introduced to solve the degeneracy of the equation $s\l+w \o+[\o,\l]=0$
 and $s\bar c+w\bar \o=0$. The conservation of both ghost numbers
 $N_s$ and $N_w$ is of course most important in this determination.

As explained in \bgz, the $s$-invariance enforces the possibility of
defining observables in any given slice, the $w$-invariance expresses the
 Yang--Mills gauge symmetry of the theory. Actually, observables are
defined as the cohomology of $w$, that one can restrict to a slice,
provided no anomaly occurs. Actually,  power counting and the
requirements of locality, $s$ and $w$ invariances, (5th) time parity
symmetry  and   ghost number conservation completely determine    the local
five-dimensional action
$I$, eq.~\cgact.

\subsec{Elimination of ghosts and auxiliary fields}

	We now show that the action $I$ is physically equivalent to the
original
action \tgfix.  The argument relies on the fact that all ghosts decouple
because all ghost propagators are retarded and ghost numbers are
conserved.  Indeed all free ghost propagators such as \frprop\ are
analytic in the lower half $\o$-plane.  Consequently all the free ghost
propagators are retarded, $D_{\l \m}^{\psi
\pb}(x, t) = 0$ for $t < 0$ and likewise for the other ghost propagators.
This is a characteristic of parabolic field equations.  Since every ghost
propagator is retarded,  {\it all closed ghost loops vanish}.\foot{There
are ghost tadpole diagrams that we neglect.  They serve only to cancel
other tadpole diagrams, and vanish with dimensional regularization.}  This
property is essential to the 5-dimensional formulation of physical
4-dimensional gauge theories. Moreover we may start at any ghost line with
non-zero $N_w$ in a Feynman diagram and follow the conserved ghost charge
$N_w$ into the future where it becomes an external $N_w$-ghost line.
Therefore each diagram with  no external $N_w$-ghosts has no internal
$N_w$-ghost line either. Consequently integration over the $N_w$-ghosts
results simply in the suppression of these ghosts in the action.  This
argument remains valid despite the presence of triple ghost vertices such
as $\bar{\m} [\l,\psi_5]$ in the action  $I$.   A similar argument also
allows us to separately integrate out the fields of the last quartet
$(\o, \phi, \bar{\phi}, \bar{\o})$.  The action \tgfix\ results.  This
also shows that the expectation-values of physical observables is
independent of the parameter $a'$ since it does not appear in \tgfix.

	 The above argument also holds for integration over ghost fields with
$N_s \neq 0$ but $N_w = 0$.  As a result, for
computing correlation functions for which all external
lines have ghost number $N_s = 0$, but possibly $N_w \neq 0$,
we may suppress all ghost fields with ghost number $N_s \neq 0$
in the action \cgact, which gives
\eqn\actw{\eqalign{
I_{\rm w}(a) = \int d^5x \Big( \pi_\m^\star(F_{5\m}
+ { {\d S \ \ } \over {\d A_\m} } + \pi_\m^\star)
- w[\bar\l(aA_5 - \p_\m A_\m)]  \Big).
}}
Here we have written
$-D_\l F_{\l\m} = { {\d S \ } \over {\d A_\m} }$, and used the
$w$-covariant field variable $\pi_\m^\star$ defined in \covb.
We will use this action in Appendix D to show invariance of
physical observables under inversion of the 5th time.  The
expectation-value of physical observables is independent of the gauge
parameter $a$ because they are in the cohomology of $w$, but the
gauge-parameter $a$ appears only in the
$w$-exact term in the actions \cgact\ or \actw.

	It also follows that the set of correlation functions
with no external ghost lines -- and this includes all physical correlation
functions -- contains no internal ghost lines either.  Consequently this
set of correlation functions is described by the action \cgact\ in which
all ghost fields are suppressed namely the reduced action
\eqn\redact{\eqalign{
I_{\rm red}
=  \int dt \ d^4x  \ \pi_\m \Big(\p_t A_\m - a^{-1}D_\m \p_\l A_\l
- D_\l F_{\l\m} + \pi_\m ).
}}
If we also
integrate over the $\pi_\m$ field (it is purely imaginary) we obtain the
completely reduced action
\eqn\credact{\eqalign{
I'_{\rm red} = - (1/4) \int dt \ d^4x \ \Big(\p_t A_\m - a^{-1}D_\m \p_\l
A_\l - D_\l F_{\l\m} \Big)^2, }}
and positive Euclidean weight
$\exp(I'_{\rm red})$ (with $I'_{\rm red} < 0$).
Upon taking the square in the last expression, the
cross-terms involving $-D_\l F_{\l\m} = {{\d S} \over {\d A_\m} }$ are
exact derivatives.  Indeed by the gauge-invariance of $S$ we have
$\int d^4 x \ \p_\l A_\l D_\m {{\d S} \over {\d A_\m} } = 0$
and also
$\int d^4 x \ \p_t A_\m {{\d S} \over {\d A_\m} } = \dot{S}$.  Thus we
may define
\eqn\totact{\eqalign{
I'_{\rm tot} & \equiv - \int dt (1/2)\dot{S} +  I'_{\rm red}   \cr
I'_{\rm tot} & = - (1/4)\int dt \ d^4x
\ [ \ (\p_t A_\m - a^{-1}D_\m \p_\l A_\l)^2 \  +  \ (D_\l F_{\l\m})^2 \ ],
}}
which is a sum of squares. When this weight is used in the path integral
over ${\cal D} A_{x,t}$ the difficulties that  Faddeev--Popov
distribution in 4 dimensions encounters at the non-perturbative level are
avoided.

\newsec{Confinement and the Gribov region}

	We have seen that use of the fifth dimension avoids the formal
difficulties of gauge-fixing in 4 dimensions that is problematical at
the non-perturbative level.  We shall now show explicitly how the
resulting local 5-dimensional action in fact concentrates the weight in
or near the Gribov region.  This can only be achieved in the
4-dimensional formulation by topological identification of boundary of
the fundamental region, which requires difficult non-perturbative
calculations \cutkosky\ and \vanbaal.  Here we recall and sharpen the
discussion of \bgz\ of the limit $a \to 0$ of the path integral \totact.
(Recall that the mean values of observables are independent of the choice
the gauge parameter $a$.)

	Upon rescaling the time according to $t \to a
t$, we obtain
\eqn\rescact{\eqalign{
I'_{\rm tot}  = - (1/4)\int dt \ d^4x
\ [ \ a^{-1} (\p_t A_\m - D_\m \p_\l A_\l)^2 \  +  a \ (D_\l F_{\l\m})^2 \
], }}
In the limit $a \to 0$, the path integral over $A_\mu(x,t)$ gets
concentrated near where the condition
$F_{5\mu} = \p_t A_\m - D_\m \p_\l A_\l = 0$, is satisfied, namely
near configurations that satisfy the flow equation
\eqn\fixed{\eqalign{
\partial _t A_\m = D_\m  \partial_\l A_\l.
}}
We now make a global analysis of this flow.  The velocity field
$D_\m   \partial_\l A_\l$ is an infinitesimal gauge transformation,
with generator $\o = \p_\l A_\l$, so the flow at each point
$A = A_\m(x)$ is tangent to the gauge orbit through $A$.  We
assert that the flow \fixed\ is, at each point $A$, in the direction
of steepest descent of the ``minimizing" functional,
\eqn\minfunc{\eqalign{
{\cal F}_A[g] = {\cal F}_{{^g}A}[1] \equiv ||{^g}A||^2,
}}
defined on the gauge orbit through $A$. Here
$||A||^2 \equiv \int d^4x |A_\m |^2$
is the 4-dimensional Hilbert-norm, and
${^g}A_\m = g^{-1} A_\m g + g^{-1}\p_\m g$ is the gauge-transform
of~$A_\m$.
To prove the assertion, consider the variation of the functional
${\cal F}_A[1] \equiv ||A||^2$
under an arbitrary infinitesimal gauge
transformation, $\d A_\m = D_\m (A)\o$,
\eqn\stdesc{\eqalign{
\d ||A||^2 = 2(A_\m,\d A_\m) = 2(A_\m,D_\m \o) = 2(A_\m,\p_\m \o)
= - 2(\p_\m A_\m, \o).
}}
Thus the direction of steepest descent of $||A||^2$, restricted to
directions tangent to the gauge orbit at $A$, is given by
the generator $\o = \p_\m A_\m$, which is what we wished to establish.
Starting from an arbitrary configuration, the flow \fixed,
$||A||^2$ decreases monotonically, $\p_t||A||^2 = -2 ||\p_\m A_\m||^2$.
Because $||A||^2$ is bounded below, we have
$\lim_{t \to \infty} ||\p_\m A_\m||^2 = 0$. This implies
$\lim_{T \to \infty} T^{-1} \int_0^T dt ||\p_\m A_\m||^2 = 0$,
or, in words, over an infinite time interval the time average of
$||\p_\m A_\m||^2$ vanishes.  Non-zero values of $\p_\m A_\m$ are mere
transients, and under the flow \fixed, during an infinite time
interval, the weight is entirely concentrated on transverse configurations
$A^{\rm tr}$ that satisfy Landau-gauge condition
$\p_\l A_\l^{\rm tr} = 0$.

	By eq. \stdesc, these are the stationary points of the minimizing
functional $F_A[g] = ||{^g}A||^2$ at $g = 1$.  Stationary points may be
either minima or saddle-points, according as the second variation of the
minimizing functional $\d^2 F[A]$, restricted to directions
$\d A_\m = D_\m (A) \o$ tangent to the gauge orbit
through $A$, is positive for all $\o$ or not.  From
\stdesc, it is given by
\eqn\hessian{\eqalign{
\d^2 F[A] = - 2 \d (\o,\p_\m A_\m )
= - 2(\o, \p_\m D_\m(A) \o).
}}
Only exceptional configurations flow to saddle-points, where the
equilibrium under the ``force" $D_\m \p_\l A_\l$ is
unstable.   (For if one moves off a point of unstable equilibrium, then in
general one picks an unstable component of the force.) Non-exceptional
configurations flow to minima, which are stable attractors.
We conclude that under the flow \fixed, and for an infinite-time
interval, the weight is entirely concentrated on configurations that
satisfy two conditions: (i) they are transverse $\p_\l A_\l = 0$
and (ii) the Faddeev-Popov operator is positive
$ - \p_\m D_\m(A) > 0$. These two conditions define the Gribov
region.  Thus in the limit $a \to 0$, the weight corresponding to the
partition function \rescact\ gets concentrated inside the Gribov region.
For small positive values of the gauge parameter $a$, the weight is smeared
out on each gauge orbit but concentrated near the Gribov region.

	The first condition, transversality, is a standard gauge-fixing
condition of the 4-dimensional formulation.  However the second condition,
the positivity of the Faddeev-Popov operator, is not achievable by a
local 4-dimensional action.  It was shown in \vanish\ that the gluon
propagator $D(k)$ vanishes at $k = 0$ for a probability distribution
concentrated in the Gribov region.  As was discussed in the Introduction,
this excludes the possibility of a pole at $k^2 = 0$ which corresponds to
a physical massless gluon.  Since poles of the propagator are
independent of the gauge parameter $a$ by virtue of the Nielsen
identities \nielsen, this conclusion holds for all values of $a$.  Absence
of a massless gluon pole is an important first step toward proving
confinement.

\newsec{Higgs phase}

	We now show how the above considerations may be extended to the
Georgi--Glashow or the standard model.  For simplicity we take the
Georgi-Glashow model, with classical 4-dimensional Euclidean action
\eqn\higgseu{\eqalign{
S = \int d^4x [ (1/4)F_{\m \n}^2  + (1/2)(D_\m \phi)^2
      + (1/4) \l (\phi^2 - v^2)^2],
}}
where $\phi = (\phi^a) = \vec{\phi}$ is in the adjoint representation of
the SU(2) group.
 We shall show how the 5-dimensional formulation
allows a global analysis of gauge fixing in the Higgs phase,
as in the pure gauge case.  This is especially important because in this
model, the perturbative and the exact spectrum do not agree.  Indeed
as is well known, a semi-classical analysis results in spontaneous
breaking of the SU(2) symmetry with a massless ``photon"
associated with the unbroken U(1) gauge symmetry, as is verified in the
present context in Appendix E.  On the other hand it has been shown that in
the Georgi--Glashow model in 3 dimensions that the so-called ``broken" and
``unbroken" phases are in fact continuously connected and moreover that no
massless photon exists due to condensation of monopoles \thooft,
\polyakov, \ostseil, \fradshen, \banksrab.  It is interesting to note that
the last statement agrees with the conclusion of the previous section that
excludes a massless gluon pole.  This motivates us to examine the
consequences of the non-perturbative gauge-fixing of the 5-dimensional
formulation when it is adapted to the case of the Georgi-Glashow model.
Moreover a consistent definition of a physical particle that is not a
singlet under a local gauge group is also difficult in the Higgs phase.

	The 5-dimensional action for the Georgi-Glashow model that replaces
\credact\ is given by
\eqn\higgsract{\eqalign{
I'_{\rm red}  = - (1/4)\int dt d^4x
\Big[ \Big(F_{5 \m} + {{\d S \ }\over {\d A_\m}}\Big)^2 +
\Big(D_5 \phi + {{\d S}\over {\d \phi}}\Big)^2\Big],
}}
where
\eqn\giggsract{\eqalign{
{{\d S \ }\over {\d A_\m}}  = - D_\l F_{\l \m} + [\phi, D_\m \phi];
\ \ \ \
{{\d S}\over {\d \phi}}  = - D_\m^2 \phi + \l (\phi^2 - v^2) \phi.
}}
The second term in this action corresponds to the Langevin equation for
$\phi$
\eqn\langevin{\eqalign{
\p_5 \phi = - [A_5, \phi] - {{\d S}\over {\d \phi}} + {\rm noise}.
}}
Here the term $- [A_5, \phi]$ acts as a restoring ``force" tangent to the
gauge orbit through $\phi$.

 Their remains to specify $A_5$ appropriate to this model.  If the gauge
choice discussed in the preceding section namely, $A_5 = a^{-1}\p_\m A_\m$
in the limit $a \to 0$, is well-defined in the present case, then the
conclusion of the previous section follows also in the Higgs phase namely
that a massless ``photon'' is excluded.  In this context it is helpful to
consider a more general class of gauges defined by the more general
 minimizing functional on the gauge orbit defined by
\eqn\higgsmf{\eqalign{
{\cal F}_{A,\phi}[g] =  {\cal F}_{{^g}A,{^g}\phi}[1]
= \int d^4x \ \big[ \ (2a)^{-1}|{^g}A|^2  - M \hat{n}\cdot {^g}\phi \
\big],
}}
where $t^a({^g}\phi)^a = g^{-1} t^a\phi^a g$.  Here the gauge parameters
are $a > 0$, $M > 0$, and the direction $\hat{n}$.  Because
$|{^g}\phi| = |\phi|$, this functional on the gauge orbit
is bounded below by $ - M \ \int d^4x  \ |\phi|$  which, for given $\phi$,
is finite for a finite Euclidean volume.  Clearly, for $M>0$
this minimizing functional favors configurations $\phi$ that are
aligned along $\hat{n}$, whereas $M = 0$ is a less complete gauge fixing.
This minimizing functional offers new possibilities that could be of
interest in the context of numerical gauge-fixing in simulations of
lattice gauge theory.

	We now analyse the gauge fixing associated with this minimizing
functional.  Under the infinitesimal gauge transformation,
$\d A_\m = D_\m \o$ and $\d \phi = [\phi, \o] $,
its first and second variations are given by
\eqn\higgsvar{\eqalign{
\d {\cal F}_{A, \phi}[1] & =
- \Big(\o, a^{-1}\p_\m A_\m + M [\hat{n}, \phi] \Big)
\cr
\d^2 {\cal F}_{A, \phi}[1] & =
- \Big(\o, a^{-1}\p_\m D_\m(A) \o + M [\hat{n}, [\phi, \o]]\Big). }}
For $A_5$ we choose the direction of steepest descent of
${\cal F}_{A, \phi}[1]$, restricted to directions tangent to the gauge
orbit,
\eqn\higgsgf{\eqalign{
\vec{A_5} =  a^{-1}\p_\m \vec{A}_\m
+ M \hat{n} \times \vec{\phi}, }}
where $\vec{a} \times \vec{b} \equiv [a, b]$.  A semi-classical analysis
of the action \higgsract\ with $A_5$ given in \higgsgf, is presented in
Appendix E, which gives the standard semi-classical result namely a pair
of charged massive gauge particles and a massless photon associated with
the unbroken U(1) symmetry.

	For the non-pertubative analysis we consider the gauge defined by
\higgsgf\ for large (positive) values of the gauge parameters $a^{-1}$
and~$M$.  We scale $M = a^{-1}M'$, and take $a$ to be arbitrarily
small.  The argument of the preceding section  may be used, with the
conclusion that in the limit $a \to 0$, the probability gets concentrated
near the minima of the minimizing functional ${\cal F}_{A,\phi}[g]$, eq.
\higgsmf, namely (i) where ${\cal F}_{A,\phi}[g]$ is stationary,
\eqn\higgsgge{\eqalign{
\p_\m \vec{A}_\m + M' \hat{n} \times \vec{\phi} = 0,
}}
and {\it in
addition} (ii) where its second variation is positive namely, by
\higgsvar,
\eqn\higgsfp{\eqalign{
\Big( \o, - \p_\m D_\m(A)\o - M' \hat{n} \times (\vec{\phi} \times \o
)\Big)  \ \geq 0     \ \ \ \ {\rm for \ all} \  \o.
}}
The second condition, which expresses the positivity of the relevant
Faddeev-Popov operator, is a new, non-perturbative condition, not
available in the 4-dimensional formulation, that expresses the
restriction to the Gribov region appropriate to this gauge fixing.
We note that both conditions are linear in the fields $A_\m$ and $\phi$.
As a result the Gribov region is convex in $A$-$\phi$ space:  if
$(A^{(i)},\phi^{(i)})$  lie in the Gribov region for $i = 1, 2$, then
$(A, \phi)$ also lies in the
Gribov region for $A = \a A^{(1)} + \b A^{(2)}$ and
$\phi = \a \phi^{(1)} + \b \phi^{(2)}$,
where $\a > 0 $ and $\b = 1-\a > 0$.

	Upon taking the vacuum expectation value of \higgsgge, one
obtains
\eqn\higgsir{\eqalign{
M' \ \hat{n} \times \langle \vec{\phi} \rangle  = 0,
}}
so for $M'$ finite, the Higgs field $\vec{\phi}$ cannot acquire a vacuum
expectation-value in the direction perpendicular to the gauge parameter
$\hat{n}$, whereas no such restriction holds at $M' = 0$.  Indeed
without an $\hat n$ dependence of the gauge-fixing drift force \langevin,
there are random walks of the Higgs field in the flat valley.  This
suggests that $M' = 0$ or $M = 0$ may be a singular point where the
gauge is not well defined.  However if the gauge $M' = 0$ is well-defined,
then the Landau-gauge condition $\partial_\mu A_\m = 0$ holds here as in
the previous section which, as we have seen, excludes a massless gauge
particle in agreement with Polyakov's conclusion \polyakov.  (We again
recall that the position of poles in a propagator is independent of the
gauge parameters by virtue of the Nielsen identities.)  Note that the
Landau gauge is ill-defined in 4-dimensional perturbative calculations
because it induces spurious double poles in the propagator of Goldstone
bosons.

	In terms of the shifted Higgs
field,  $\vec{\phi} = \vec{v} + \vec{\phi}'$, where
$\langle\vec{\phi}\rangle = \vec{v}$, the positivity condition reads
\eqn\higgsfpa{\eqalign{
\Big( \o, - \p_\m D_\m(A)\o - M' \hat{n} \times (\vec{\phi}' \times \o)
+  M'v( \o - \hat{n} \hat{n} \cdot \o) \Big)  \ \geq 0     \ \ \ \ {\rm
for \ all} \  \o.
}}
The  neutral component of $\vec{A}_\m$
(i. e. along the $\hat{n}$-direction) is restricted only by the components
of $\o$ that are perpendicular to $\hat{n}$.  So for the neutral
component the positivity condition is expressed by
\eqn\higgsfpb{\eqalign{
\Big( \o, - \p_\m D_\m(A)\o - M' \hat{n} \times (\vec{\phi}' \times \o)
+  M'v \  \o \Big)  \ \geq 0     \ \ \ \  \o \perp \hat{n}.
}}
The last term is strictly positive for $M' > 0$, so the
restriction on the neutral component is qualitatively
weaker than in the $M' = 0$ case and may not be incompatible with a
massless gauge particle.  The gauge condition now depends on the
parameters of the Higgs sector of the model, such as $v$, as well as on the
gauge parameters, such as $M$, $a$ and $\hat{n}$.  Depending on the
values of these parameters, the
restriction to the Gribov region may give valuable information about the
phase such as the position of poles of propagators.
This information could be obtained from calculation of propagators by
numerical simulation and minimization of a lattice analog of the minimizing
functional \higgsmf\ such as
\eqn\higgsmfa{\eqalign{
{\cal F}_{U,\phi}[g] =  {\cal F}_{{^g}U,{^g}\phi}[1]
= \sum_x \ \big[ - 2a^{-1}
\sum_\m   {\rm Re \ tr} {^g}U_{x,\m}
  - M \hat{n}\cdot {^g}\phi_x \
\big],
}}
in the notation of \zbgr.

\newsec{Conclusion}

	As an alternative to the geometric method presented in \bgz, in the
present article we derived the bulk or stochastic quantization of a gauge
field in a series of intuitive steps.   The starting point is the bulk
quantization of fields of non-gauge type presented in the preceding
article \bulkq.  Whereas the standard Faddeev-Popov method relies on
{\it gauge-fixing} that is subject to the problem of Gribov copies, in
the step-by-step construction, gauge-fixing is replaced by an
$A$-dependent {\it gauge transformation} whose Jacobian
is an infinite constant that cancels the divergent volume of
the gauge group.  We have shown the perturbative equivalence of the 4- and
5-dimensional formulations of gauge theories by showing that in Landau
gauge the Schwinger-Dyson equations of the 4-dimensional theory hold on a
time slice of the 5-dimensional theory.  We refer to the preceding article
\bulkq\ for a discussion of the S-matrix that could be formally applied to
the case of gauge theories treated perturbatively.

	As for physical applications, we have shown that in the limiting
case of
large gauge parameters, bulk quantization of gauge fields automatically
restricts the probability to the interior of the Gribov region in the
context of a local, renormalizable theory.  For the case of a pure gauge
theory, this excludes the existence of physical massless gauge quantum, a
first step toward proving confinement.  A new result is a minimizing
functional \higgsmfa\ which is appropriate to global gauge fixing in the
presence of coupling to a Higgs field, for which we have found the
corresponding Gribov region.  The lattice analog of this minimizing
functional \higgsmfa\ may be used for numerical gauge fixing in simulations
of lattice gauge theory.

	In this connection we wish to emphasize that lattice discretization
of the
5-dimensional theory \zbgr\ offers distinct computational possibilities
from Monte Carlo simulations of the lattice discretization of the
4-dimensional theory using detailed balance.   Discretization of the
5-dimensional theory corresponds to simulation of the Langevin equation
with time-step $\e \sim a^2$, and it is sufficient that they agree in the
limit $a \to 0$ \batrouni, \davies.  These studies and others \fukugita,
\kronfeld\ have addressed the question of whether the 4- and 5-dimensional
discretizations of gauge theories fall into the same universality class
and have shown that they do, to first order in $\e$.  The present
approach, in which the renormalizability of the local 5-dimensional
formulation of a gauge theory is assured \bgz, provides an affirmative
answer to this question to all orders.

\vskip .5cm
{\centerline{\bf Acknowledgments}}

It is a pleasure to thank P.~A.~Grassi for valuable discussions.
The research of Daniel Zwanziger was partially supported by the National
Science Foundation under grant PHY-9900769.
\vskip .5cm

\appendix A{Continuation of Sec 2: Steps 4 and 5}

\subsec{Step 4: BRST implementation of the 5-dimensional gauge invariance}

The most expedient way to preserve the 5-dimensional gauge
symmetry
\gtrans\ and \sgtrans\ is to encode it in a second operator BRST
operator $w$ that generates an infinitesimal gauge transformation
in the usual way,
\eqn\wact{\eqalign{
 w A_\m & = D_\m \l \ \ \ \ \ \ \ \ w A_5 = D_5 \l \cr w \l & = -
\demi [\l, \l],  }}
and satisfies $w^2 = 0$.  The new Fermi ghost field $\l$ reminds us of
the familiar Faddeev-Popov ghost.

	We require that the two BRST operators $s$ and $w$ be algebraically
consistent in the sense that
\eqn\cnsst{\eqalign{
 s^2 = w^2 = sw + ws = 0 }} holds.  We also want to construct an action
that is both $s$- and $w$-invariant and that is physically equivalent to
\tgfix.  We shall modify the
action $\hat{I}_{\rm YM}$ by additional ghost terms that involve
additional ghost fields, and we shall show in the following
section that the action we obtain is physically equivalent to
$\hat{I}_{\rm YM}$.

	The principle that we use to construct a consistent algebra for $s$
and $w$ is that $s$ should act trivially in the sense that it acts on
an elementary field to produce an elementary field rather than a
composite.  Accordingly we put $\l$ into a new quartet
$(\l, \m, \bar{\m},\bar{\l})$ within which $s$ acts trivially as before,
and the action of $s$ on all fields is given in eq. \sonall.

Here $\m$ is a new scalar Bose ghost field
that is the topological ghost of the ghost $\l$, and
$\bar{\l}$ and $\bar{\m}$ are the corresponding anti-ghosts.  In
general we use the ``bar" to indicate the anti-ghost of the
corresponding ghost, which is also its canonical momentum density,
except for  $\mb$, which is the anti-ghost of $\psi_5$, but is not its
canonical momentum density in the sense that it enforces a constraint.
The last line is a new quartet $(\o, \phi, \bar{\phi}, \bar{\o})$ that
will be introduced below.

	As regards algebraic consistency, we may assign as convenient the
$w$-transform of any of the above elementary fields that is not an
$s$-transform, provided only that it is consistent with $w^2 = 0$.  The
action of $w$ on any of the above elementary fields that is an
$s$-transform is then determined by the consistency condition
$sw + ws = 0$.

	We have already stated the $w$-transforms of the fields
$A_\n$, $A_5$ and $\l$.  Accordingly the $w$-transforms of their
$s$-transforms $\psi_\n = sA_\n$, $\psi_5 = sA_5$,
and $\m = s\l$ are determined by algebraic consistency, namely,
\eqn\fwonp{\eqalign{
 w \psi_\n & = wsA_\n = - swA_\n = -sD_\n \l
                = - [\l , \psi_\n] - D_\n \m,   \cr
 w \psi_5 & = wsA_5 = - swA_5 = -sD_5 \l
                = - [\l , \psi_5] - D_5 \m,    \cr
 w \m & = ws\l = - sw\l = \demi s [\l, \l]
= \demi ([\m, \l] - [\l, \m]) = - [\l, \m].
}}
We now turn to the anti-ghosts $\pb_\n$, $\mb$ and $\bar{\m}$ that are not
the $s$-transforms of anything.  It will be useful for the construction of
an $s$ and $w$-invariant action to assign them the transformation law
\eqn\wonpb{\eqalign{
w \pbi_\n = - [\l, \pbi_\n];
\ \ \ \ \ \ \ \ w \bar{\m} = \mb;
  \ \ \  \ \ \ \ \ w \mb = 0   }}
which is consistent with $w^2 = 0$.  The $w$-transforms of their
$s$-transforms $\pi_\n = s \pbi_\n$, $\bar{\l} = s \bar{\m}$, and
$l = s \bar{m}$ are determined by algebraic consistency,
\eqn\fwonb{\eqalign{
w \pi_\n & = ws\pbi_\n = - sw\pbi_\n = s [\l, \pbi_\n]
    = - [\l, \pi_\n] + [\m, \pbi_\n];   \cr
w \bar\l & = w s \bar{\m} = - s w \bar{\m} = - s \bar{m} = - l;  \cr
w l & = w s \bar{m} = - s w \bar{m} = 0.
}}
One may verify that $w^2 = 0$ is maintained.  We have now determined the
action of $w$ on all quartets appearing in \sonall\ except the last one,
which will be determined below, with the result given in \wonall.

	Because $w$ generates an infinitesimal gauge transformation on $A_\m$
and $A_5$,  the fields $F_{5\m}$ and $D_\l F_{\l\m}$
transform gauge covariantly,
$w F_{5\m} = -[\l, F_{5\m}]$ and $w D_\m F_{\m\n} = -[\l, D_\m F_{\m\n}]$.
The anti-ghost field $\pbi_\n$ was chosen to also transform
gauge-covariantly $w \pbi_\n = - [\l, \pbi_\n]$, so the first term
of the $s$-exact action \tgfix,
\eqn\ffact{\eqalign{
  I_F \equiv \int d^5x \ s[ \ \pbi_\m (F_{5\m} - D_\l F_{\l\m} ) \ ],
}}
\eqn\xffact{\eqalign{
I_F
=  \int dt \ d^4x \ \Big[ & \ \pi_\m \Big(F_{5\m} - D_\l F_{\l\m}  \Big)
 - \ \pbi_\m \Big(D_5 \psi_\m - D_\m \psi_5\cr
 & \ \ \ \ \ \ \ \ \ \ \ \ \ \ \ \ \ \ \ \ \ \ \ \ \ \ \ \ \ \ \ \  - D_\l
(D_\l \psi_\m -
D_\m\psi _\l)
   - [\psi_\l, F_{\l\m}] \Big)  \ \Big],
}}
is $w$-invariant,  $wI_F = 0$, where $I_F$ is written explicitly below.
In fact it is in the cohomology of $w$, because it is not $w$-exact, $I_F
\neq wX$.

	To impose the gauge conditions $aA_5 = \p_\m A_\m$
and $a\psi_5 = \p_\m \psi_\m$ in a way which is
consistent with both $s$ and $w$ invariance, we take instead of
\gfixa\ the gauge-fixing action,
\eqn\gfixb{\eqalign{
  I_{\rm gf} \equiv \int d^5x \ sw [ \ \bar{\m}(a A_5 - \p_\n A_\n) \ ],
}}
\eqn\zgfix{\eqalign{
  I_{\rm gf} = \int d^5x \ \Big[ \ & l(a A_5 - \p_\n A_\n) -
\mb(a \psi_5 - \p_\n \psi_\n) + \bar{\l}(a D_5 \l - \p_\n D_\n \l)
\cr
& + \bar{\m}\Big( \ a D_5 \m - \p_\n D_\n \m + a[\psi_5, \l] -
\p_\n [\psi_\n, \l] \ \Big) \ \Big];
}}
that is both $s$- and $w$-exact. The
first two terms agree with the action $\hat{I}_{\rm gf}$, eq. \gfixa,
which imposes the desired constraints.  With $a > 0$, the remaining terms
in the action provide parabolic field equations for the new ghosts $\l$
and~$\m$.

\subsec{Step 5: Construction of $w$-covariant fields}

	Having chosen the transformation law of $\pbi_\n$ to be covariant under
$w$, it is inevitable that $\pi_\n = s \pbi_\n$ does not transform
covariantly under $w$, as one sees from \fwonb.  As a result, the last
term of \gcgact, $\int d^5x \ s(\pb_\n \pi_\n)$ is {\it not}
$w$-invariant.  One way to overcome this
difficulty is to replace $\pi_\n$ by $\pi^\star$ defined in \covb.
Here $\o$ is some Fermi-ghost
field whose transformation law under $w$ must be such that
$\pi^\star_\n$ is gauge covariant,
\eqn\woncb{\eqalign{
w \pi^\star_\n = - [\l, \pi^\star_\n], }}
so that the action,
\eqn\bact{\eqalign{
  I_\pi \equiv \int d^5x \ s( \ \pbi_\m \pi^\star_\m \ )
}}
\eqn\bpact{\eqalign{
  I_\pi   =  \int  d^5x \ s\Big(  \ \pbi_\m (\pi_\m + [\o, \pbi_\m])
\ \Big)    =
    \int  d^5x \ \Big(  \ \pi_\m \pi_\m + 2 \pi_\m [\pbi_\m, \o]
 + [\pbi_\m, \pbi_\m]\f   \ \Big),
}}
is both $s$-exact and $w$-invariant, $wI_\pi =
0$.  One easily verifies that $\pi^\star_\n$ does transform
gauge-covariantly, provided that
$\o$ satisfies the transformation law $w\o = - [\l, \o] - \m$ that appears
in the last line of \wonall.  It is consistent with $w^2\o = 0$.

	None of the fields we have introduced so far have this transformation
law, so we take $\o$ to be a new elementary field.
It might be called an ``adjuster" field because it allows us to
``adjust" $\pi_\n$ to make a new field that transforms covariantly.  To
maintain the trivial action of the $s$-operator, we take the new field
$\o$ to be part of a new quartet $(\o, \f, \bar{\f},
\bar{\omega})$ within which $s$ acts as shown in the last line of
\sonall.  The action of $w$ on $\f$ is determined by
\eqn\fwonf{\eqalign{
w\f  = ws\o = -sw\o = s([\l, \o] + \m) = [s\l, \o] - [\l, s\o]
= - [\l, \f] + [\m, \o].
}}
We also assign $\bar{\phi}$ to be $w$-covariant, which also determines
$w\bar{\o}$.  These relations are shown in \wonall.   The action \bact\
is $s$-exact and in the cohomology of $w$, $wI_\pi = 0$.

	We require an action $I_\o$ to provide equations of motion
for the new quartet $(\o, \f, \bar{\f},
\bar{\o})$ that should also be $s$-exact and $w$-invariant.  To find it,
observe that the field $\o$ also allows us to ``adjust" the ghost
fields $\psi_\m$ and $\psi_5$, so the adjusted fields \covpsi\
are $w$-covariant,
\eqn\covwonp{\eqalign{
w \psi^\star_\m = - [\l, \psi^\star_\m], \ \ \ \ \ \ w\psi^\star_5
= - [\l, \psi^\star_5],
}}
as is easily verified.  The combination
$D_\n \psi^\star_\n$ is also gauge covariant.  Because $\bar{\f}$,
transforms $w$-covariantly,
$w\bar{\f} = - [\l, \bar{\f}]$, the $s$-exact action,
\eqn\cact{\eqalign{
  I_\o & \equiv \int d^5x \ s[ \ \bar{\f}( a' \psi^\star_5 - D_\m
\psi^\star_\m) \ ]  \cr
 & = \int d^5x \ s\{ \ \bar{\f}[ a' (\psi_5 - D_5 \o)
- D_\m (\psi_\m - D_\m \o)] \ \}     \cr
 & =  \int d^5x \ \Big[ \ \bar{\o}
 \Big(- (a'D_5 - D_\n D_\n) \o
+ a' \psi_5 -  D_\n \psi_\n \Big)
 +   \  \bar{\f}
\Big( - (a'D_5 - D_\n D_\n) \f  \cr
& \ \ \ \ \ \ \ \ \ \ \ \ \ \ \ \ -a'[\psi_5,\o] + [\psi_\n, D_\n \o] +
D_\n [\psi_\n, \o] - [\psi_\n, \psi_\n] \Big) \ \Big],
}}
is $w$-invariant, $wI_\o = 0$.  It provides parabolic equations of motion
for $\o$ and $\f$, as long as the otherwise arbitrary parameter $a'$ is
positive, $a' > 0$.  The total action
\eqn\cgactxp{\eqalign{
  I & \equiv I_F + I_\pi + I_\o + I_{\rm gf}
}}
is given in \cgact.  This completes the step-by-step construction of the
TQFT for a gauge theory.

\appendix B{Jacobian of gauge transformation}

As announced in section 2, we must check that  the Jacobian $J$ of the
transformation $A
\to {^gA}$ is $A$-independent, $J = {\rm const}$. It is sufficient to do
this for the infinitesimal gauge transformation that changes the gauge
parameter $\a \equiv a^{-1}$ by an infinitesimal amount
$\e$.  Let us determine the infinitesimal gauge transformation $\d A_\m =
D_\m(A)\o$ that achieves this.  Assume that $A_5 = \a \p_\m A_\m$, and
that $A'_5 = (\a + \e) \p_\m A'_\m$, where $A'_5 = A_5 + D_5 \o$ and
$A'_\m = A_\m + D_\m \o$, and $\o = O(\e)$.  To first order in $\e$, the
last two equations give the condition on $\o$,
\eqn\rlt{\eqalign{
D_5\o - \a \p_\m D_\m \o & = \e \p_\m A_\m  \cr
\p_5\o - \a D_\m \p_\m \o & = \e \p_\m A_\m.
}}
This is a linear, inhomogeneous, parabolic equation for $\o$.  It
has the unique solution
\eqn\solom{\eqalign{
\o^a(x,t, A) =
\e \int_{-\infty}^t du \ d^4y \ G^{ab}(x,t; y,u;A) \ \p_\l
A_\l^b(y,u), }} where $G$ is the Green's function
defined by
\eqn\grfn{\eqalign{
(\p_5 - \a D_\m \p_\m)G(x,t; \ y,u) = \d(x-y) \d(t-u).
}}

	We now calculate the Jacobian of the infinitesimal transformation
$A_\m' = A_\m + D_\m(A)\o$.  For an infinitesimal transformation
with discrete variables, $x_i' = x_i + \e f_i(x)$, say, the Jacobian
is given by $J = 1 + \e \p f_i/\p x_i$, where the second term is a
divergence.  Thus the Jacobian which we must evaluate is given by $J =
1 + K$, where $K$ is the functional trace,
\eqn\trace{\eqalign{K =  \int  dt \ d^4x
{ {\d (D_\m \o)^a(x,t)} \over {\d A_\m^a(y,u)} }|_{y=x,u =t} . }}
To evaluate the functional derivative, consider the variation induced in
$(D_\m \o)^a(x,t)$ by an infinitesimal variation $\d A_\m^a(y,u)$
\eqn\var{\eqalign{
\d (D_\m^{ac} \o^c) =  D_\m^{ac} \d \o^c
   + f^{abc} \d A_\m^b \o^c.}}
Because of the anti-symmetry of the structure constants, the second term
does not contribute to the trace, and it is sufficient to consider the
variation $D_\m^{ac} \d \o^c$.  With $\o$ given in \solom, we have
\eqn\varom{\eqalign{
\d \o^a(x,t) =
 \e  \int_{-\infty}^t du \ d^4y [ & \ G^{ab}(x,t; y,u;A) \ \p_\l \d
A_\l^b(y,u)\cr
  & + \d G^{ab}(x,t; y,u;A) \ \p_\l A_\l^b(y,u)]. }}
We will use the following properties of $G$:
\eqn\propg{\eqalign{
G^{ab}(x,t;z,v;A) = & \ \d^{ab}G_0(x-z,t-v)     \cr
  & + \int_v^t du  \ d^4yG_0(x-y,t-u) \ \a f^{acd} A_\l^c(y,u) \p_\l
      G^{db}(y,u;z,v;A), }}
\eqn\propvg{\eqalign{
\d G^{ab}(x,t;z,v;A) =
   \int_v^t du  \ d^4y \ & G^{ac}(x,t;y,u;A)    \cr
& \times \a f^{cde} \d A_\l^d(y,u)  \p_\l G^{eb}(y,u;z,v;A), }}
where $G_0(x,t)$ is the free Green function,
\eqn\freeg{\eqalign{
(\p_t - \a \p_\l \p_\l)G_0(x,t) & = \d(t) \ \d^4(x) \cr G_0(x,t) &
= \theta(t) \big({ {a} \over {4 \pi t} } \big)^2 \exp\big( - { {ax^2}
\over {4t} } \big).}}
Since we will take the trace, it is sufficient
to evaluate $\d \o(x,t)$ for variations $\d A_\m^a(y,u)$ for $u$
close to $t$, which greatly simplifies the calculation.  Indeed, for
$u$ close to $t$ we have
\eqn\aproxg{\eqalign{
G^{ab}(x,t;y,u;A) \approx \d^{ab}G_0(x-y,t-u) ,}} because the range of
the time integration in the second term of \propg\ is negligible.  For
the same reason, for variations $\d A_\m^d(y,u)$ which are non-zero
only for $u$ close to $t$, we have $\d G^{ab}(x,t;z,v;A)
\approx 0$, by eq. \propvg.  For these variations we may replace \varom\ by
its approximate
expression
\eqn\apvarom{\eqalign{
\d \o^a(x,t) \approx
 \e  \int_{-\infty}^t du \ d^4y  \ G_0(x-y,t-u) \ \p_\l \d A_\l^a(y,u). }}

	With this result, we obtain for the required variation $\d (D_\m^{ac}
\o^c) = D_\m^{ac} \d \o^c $
\eqn\rvar{\eqalign{
\d (D_\m^{ac} \o^c) = \e \ D_\m^{ac}
     \int_{-\infty}^t du \ d^4y  \ G_0(x-y,t-u) \ \p_\l \d A_\l^c(y,u).  }}
To evaluate $K$ which is the trace, eq. \trace, we need only the diagonal part
of the variation, so by the anti-symmetry of $f^{abc}$ we may replace this by
\eqn\rvar{\eqalign{
\d (D_\m^{ac} \o^c) =  - \e \ \p_\m
     \int_{-\infty}^t du \ d^4y \ G_0(x-y,t-u) \ \p_\l \d A_\l^a(y,u).
}}
The coefficient of $\d A_\l^a(y,u)$ is independent of $A$.
Consequently $K$ is independent of $A$, and thus so is the Jacobian $J
= 1 + K$.  Thus $J$ is a (divergent) constant as asserted.  The
demonstration relied heavily on the retarded properties of the Green
function of parabolic operators.

\appendix C{Equivalence of standard and bulk quantization for gauge
theories}

\def\O{{\cal O}}

	In this Appendix we shall revisit the proof that, perturbatively, the
Faddeev--Popov formulation in 4 dimensions gives the same result as the
present 5-dimensional formulation for gauge-invariant
observables. To do this we shall show that the two formulations give the
same correlation functions -- including gauge non-invariant ones -- in
the Landau-gauge limit, $a \to 0$. Because both formulations are
gauge-parameter independent for gauge-invariant quantities, this
establishes the perturbative equivalence of the two theories.  The limit
$a \to 0$ is delicate in the 5-dimensional formulation because some
propagators become elliptic instead of parabolic, and individual
Feynman diagrams may be singular at $a = 0$.  However we expect that the
correlation functions remain finite in this limit.  Technically one can
compare this limit to the one encountered when one regularizes
the singular behavior of the Coulomb gauge in 4 dimensions by a
renormalizable gauge $\xi \pa_0 A_0 +\pa_i A_i = 0$, with $\xi\to 0$
\coulomb.

	Consider a generic functional $\O = \O[A]$ of the 4-dimensional gauge
theory, not necessarily gauge invariant, whose expectation-value we wish
to compute.  It is sufficient to take $\O = \exp(J,A)$, in which case the
expectation-value $\langle\O\rangle = Z(J)$ is the generating functional
of all correlation functions.  In the perturbative 4-dimensional
formulation, $\langle \O \rangle$ is computed using the path integral
weighted by the exponential of the Faddeev--Popov action
\eqn\fpa{\eqalign{
S_{\rm FP}= \int d^4x \ [(1/4) F_{\m\n}^2
- \pa_\m\bar \y D_\m \y+\pa_\m h A_\m+M h^2].
}}
This action is invariant under the
ordinary BRST-invariance of the Faddeev-Popov theory
$ {\underline s}  S_{\rm FP} = 0$, where
${\underline s}A_\m = D_\m \y, \ {\underline s}\y = - \y^2, \
{\underline s}\bar{\y} = h, \ {\underline s}h = 0$.
From the identity,
$ 0 = \int d\F \ { {\d \ \ \  }\over {\d A_\m} }[\O \exp(-S_{\rm FP})]$,
we obtain order by order in perturbation theory, the following
Schwinger-Dyson (SD) equation:
\eqn\sd{\eqalign{
\langle  \ { {\d \O \ } \over {\d A_\mu} }
    - \O (- D_\lambda F_{\l\m} +
[\pa_\m \bar\y, \y]+\pa_\m h) \  \rangle_{\rm FP} = 0,
}}
where the mean value is computed perturbatively from the Faddeev-Popov
action \fpa.  We next write $h = {\underline s}\bar{\y}$, and use
BRST-invariance to reexpress the the last term,
\eqn\reex{\eqalign{
\langle  \ \O \ \pa_\m h(x) \ \rangle_{\rm FP}
= \langle \ \O \ {\underline s} \pa_\m  \bar{\y}(x) \ \rangle_{\rm FP}
= - \int d^4 y \ \langle \ { {\d \O \ \  } \over {\d A_\l(y) } }
(D_\l \y)(y) \ \p_\m \bar{\y}(x) \  \rangle_{\rm FP},
}}
so the SD equation reads
\eqn\sda{\eqalign{
\langle  \  { {\d \O \ } \over {\d A_\l} }(x)
+ \int d^4 y \ \p_\m \bar{\y}(x) \ \y(y)
D_\l  { {\d \O \ } \over {\d A_\l} }(y) - \O (- D_\lambda F_{\l\m} +
[\pa_\m \bar \y,\y])(x) \  \rangle_{\rm FP} = 0.
}}
Finally, we integrate out the Faddeev-Popov ghosts, and use
$\langle \bar{\y}(x) \y(y) \rangle = M^{-1}(x,y;A)$, where
$M(A) = - D_\m(A) \p_\m $ is the (Hermitian conjugate of the)
Faddeev-Popov operator, which gives
\eqn\sdb{\eqalign{
\langle  \  (I + \p M^{-1}D )_{\m \l} { {\d \O \ } \over {\d A_\l} }
- \O (- D_\lambda F_{\l\m} + \p_\m M^{-1}\  \rangle_{\rm FP}
= 0,
}}
where is last term is given explicitly by
$(\p_\m M^{-1})^a(x) \equiv f^{abc}\p_\m (M^{-1})^{bc}(x,y)|_{y=x}$.

	We now show that this equation holds in the 5-dimensional theory
by a generalization of the theory of non-gauge type discussed in \bulkq.
We start with the identity
\eqn\iden{\eqalign{
\int d\Phi { {\d } \over {\d \pi(x)}} \Big( \O[A] \exp I \Big) = 0.
}}
Here the integral is over all fields of the 5-dimensional theory, with
action \cgact, but the observable $\O[A]$ depends only on
$A_\m = A_\m(x_\l, 0)$ at $t = x_5 = 0$.  This is a generic physical
observable, and coincides wtih the observable in eq.~\sd.
From the action \cgact\ one obtains
\eqn\pieqn{\eqalign{
\langle \ \O \ (F_{5\m} - D_\lambda F_{\l\m} + 2 [\o,\pb_\mu] + 2 \pi_\m)
(x) \ \rangle_{\rm TQFT_5} = 0,
}}
where the argument $x = (x_\l,0)$ is also at $t = x_5 = 0$.

 We shall show that in the Landau gauge, $a = 0$, this equation
reduces to the form \sdb.  In Appendix C, it is proven that in this
gauge, $F_{5\m}$ is odd under time-reversal,
$F_{5\m}(x_\l, 0) \to - F_{5\m}(x_\l,0)$, whereas $\O[A]$ is even,
$\O[A] \to \O[A]$, for quantities $\O[A]$ that depend only on
$A_\m = A_\m(x_\l, 0)$ at $t = 0$.
As a result the first correlator in \pieqn\ vanishes,
$ \langle \O[A] \ F_{5\m} \rangle_{TQFT_5} = 0$.

	We next write $\pi_\m = s\pb_\m$, and use s-invariance
to rewrite the the last term,
\eqn\reexv{\eqalign{
\langle  \ \O \ \pi_\m (x) \ \rangle_{\rm FP}
= \langle \ \O \ s \pb_\m  (x) \ \rangle_{\rm
TQFT_5} = - \int d^4 y \ \langle \ { {\d \O \ \  } \over {\d A_\l(y) } }
\psi_\l(y) \ \pb_\m (x) \  \rangle_{\rm TQFT_5}.
}}
Here the $t = x_5$ component of $y = (y_\l,0)$ vanishes because
$\O[A]$ only depends upon $A$ at $t = 0$.  This gives
\eqn\sdt{\eqalign{
\langle 2\int d^4 y \ \pb_\m(x) \ \psi_\l(y)
{ {\d \O \ } \over {\d A_\l} }(y) + \O (- D_\lambda F_{\l\m} +
2[\o, \pb_\m ])(x) \  \rangle_{\rm TQFT_5} = 0.
}}

	We now evaluate the equal-time
ghost propagators that appear here in terms of the $A$-field.  Although
the action contains cubic ghost terms nevertheless, because the ghost
action is parabolic, the ghost-field propagators at equal time do not
depend on the interaction and may be evaluated exactly.  To evaluate them
we expand the action \cgact\ and obtain
\eqn\canghact{\eqalign{
I = \int d^5x \Big[ & ... - \pb_\m (D_5 \psi_\m - D_\m \psi_5 + ...)
	+  \bar\o [a' (\psi_5 - D_5 \o) + ... ]   \cr
  & + l(a A_5 - \p_\m A_\m) - \mb(a \psi_5 - \p_\m \psi_\m) + ... \Big],
}}
where we have used \xffact, \zgfix and \cact.

	We now set $a = 0$ in this expression.  The variables $A_5$ and
$\psi_5$ are no longer constrained by the gauge condition.  Integration on
$\mb$ and $\psi_5$ respectively imposes the constraints
$\p_\m \psi_\m = 0$, and $a'\bar\o = D_\m \pb_\m$.  In terms
of the remaining variables $\psi_\m^{\rm tr}$, $\o$ and $\pb_\m =
\pb_\m^{\rm tr} + \p_\m \bar\r$, the action \canghact\ becomes
\eqn\expghact{\eqalign{
I & = \int d^5x \Big[ ... - \pb_\m (\p_5 \psi_\m^{\rm tr}  + ...)
	-  D_\m \pb_\m (\p_5 \o + ... )
   - l \p_\m A_\m +  ... \Big]  \cr
 & = \int d^5x \Big[... - \pb_\m^{\rm tr} (\p_5 \psi_\m^{\rm tr}  + ...)
	-  (D_\m \pb_\m^{\rm tr} + D_\m \p_\m \bar\r) (\p_5 \o + ... )
   - l \p_\m A_\m +  ... \Big].
}}
The ghost propagators at equal time are determined by the terms in
$\p_5$ only.  In fact, for a generic parabolic action of the form
$\int dtd^4x (L_{ij}\bar\s_j \p_5 \s_i + ...)$, where $L$ is a
time-independent linear operator, the equal-time propagator is given by
$\s(x) \bar\s(y) = (1/2)L^{-1}(x,y)$.  As a result we have
$\langle \psi_\l(x) \pb_\m(y) \rangle_{\rm TQFT_5}
= (1/2)(I + D M^{-1}\p) $, and
$\langle [\o(x), \pb_\m(y)] \rangle_{\rm TQFT_5}
= (1/2)\p_\m M^{-1}(x,y)$, where quantities are defined as in eq. \sdb.
With the substitution of these values, eq. \sdt\ of the
5-dimensional theory agrees with the 4-dimensional SD equations \sdb.

	We have established that in the Landau gauge limit of the 5-dimensional
theory the SD equations of the 4-dimensional theory are satisfied.  This
shows that all correlation functions of the $A_\m$-field agree in this
gauge.  Note that if we compare \fpa\ with \expghact, we find the
interesting correspondences in Landau gauge of the 4-dimensional
Faddeev--Popov ghost with the bulk quantities:
 $\y \leftrightarrow \o$ and $\bar\y \leftrightarrow \bar\rho$.

	The proof given here is an alternative to the one displayed in
\equivstoch, still in the context of perturbation theory, which to our
knowledge was the only existing one for comparing the predictions of
both formulations. That proof relied on a definition of correlation
functions as the solution of a Fokker-Planck process that involved  a
relaxation to equilibrium and some non-local interactions, while the
proof we have just given in this paper relies on a local quantum field
theory in 5 dimensions that moreover is time-translation invariant.
We do not expect that the Faddeev--Popov measure allows one to compute
beyond perturbation theory, while, on the other hand, the 5-dimensional
formulation is expected to also hold non-perturbatively.

\appendix D{Proof of time-reversal invariance}

	Here we extend the argument of \bulkq\ to gauge theories.
Consider the
$w$-invariant action
\eqn\actwa{\eqalign{
I_{\rm tot,w}(a) \equiv I_{\rm w}(a) + \demi \int d^5x \
{ {\d S \ \ } \over {\d A_\m} }F_{5\m}.
}}
which differs from \actw\ by the second term which is an exact
derivative.   Indeed we have
$F_{5\m} = \dot{A}_\m - D_\m A_5$ and
$\dot{S} = \int d^4x { {\d S \ } \over {\d A_\m} }\dot{A}_\m$, and
moreover $D_\m { {\d S \ } \over {\d A_\m} } = 0$ by gauge invariance of
$S$. All terms in the actions \actw\ and \actwa\ are separately
$w$-invariant because both $\pi_\m^\star$ and $F_{5\m}$ are
$w$-covariant, $S$ is gauge covariant, and $w^2 = 0$.
Upon expansion, the action \actw\ at $a = 0$ reads
	\eqn\actwo{\eqalign{
I_{\rm w}(0) = \int d^5x [ \  \pi_\m^\star(F_{5\m}
+ { {\d S \ \ } \over {\d A_\m} } + \pi_\m^\star)
- \bar\l \p_\m D_\m \l
+ l \p_\m A_\m  \ ].
}}

	We shall show that $I_{\rm tot,w}(0)$ is invariant under the
time-reversal transformation
\eqn\timerev{\eqalign{
A_\m(x,t) & \to A_\m^{\rm T}(x, t) = A_\m(x, - t)  \cr
A_5(x,t) & \to A_5^{\rm T}(x, t) = - A_5(x, - t)  \cr
\pi_\m^\star(x,t) & \to \pi_\m^{\star {\rm T}}(x, t)
       = - \pi_\m^\star(x, - t) - { {\d S \ } \over {\d A_\m} }(x, -t) \cr
\l(x,t) & \to \l^{\rm T}(x, t) = \l(x, - t)  \cr
\bar\l(x,t) & \to \bar\l^{\rm T}(x, t) = \bar\l(x, - t)    \cr
l(x,t) & \to l^{\rm T}(x, t) = l(x, - t).
}}
In terms of the variables $F_{5\m}$ and
$\pi_\m' \equiv \pi_\m^\star + \demi { {\d S \ } \over {\d A_\m} }$
these transformations imply
\eqn\timerevp{\eqalign{
\pi_\m'(x,t) & \to \pi_\m'^{\rm T}(x, t)
       = - \pi_\m'(x, - t)    \cr
F_{5\m}(x,t) & \to F_{5\m}^{\rm T}(x, t)
       = - F_{5\m}(x, - t),
}}
the action $I_{\rm tot,w}(0)$ reads
	\eqn\actwp{\eqalign{
I_{\rm tot,w}(0) = \int d^5x \Big[\pi_\m' F_{5\m} + \pi_\m'^2
    - \quart \Big( { {\d S \ } \over {\d A_\m} }\Big)^2
     - \bar\l \p_\m D_\m \l + l \p_\m A_\m  \Big].
}}
This action is manifestly invariant under the above transformation.
Note that the symmetry $t \to -t$ is violated by the $w$-exact term in the
action \actw\ for $a \neq 0$.  This is a symmetry of the observables since
they are defined as the cohomology of $w$ at $t = 0$.

	We have proven that in the Landau gauge, $a = 0$, the action
$I_{\rm tot,w}(0)$ is invariant under the time reversal transformation.
This is a singular gauge in the 5-dimensional formulation.  However we
expect that the correlation functions calculated at finite $a$
have a finite limit $a \to 0$ which enjoys this symmetry.

\appendix E{Semi-classical analysis of the Higgs phase}

	We now make a semi-classical analysis of the action \higgsract, with
$A_5$ given in \higgsgf.  We shift $\vec{\phi}$ by
$\vec{\phi} = \vec{v} + \vec{\phi}'$,
where $\vec{v} \equiv v \hat{n}$ has the magnitude $v$ that appears in
the classical action \higgseu, and $\hat{n}$ is the direction that
appears in the gauge choice \higgsgf.  This gives
\eqn\lineara{\eqalign{
\vec{F}_{5\m} + {{\d S \ } \over {\d \vec{A}_\m}} =
\p_5 \vec{A}_\m & - \p_\l (\p_\l \vec{A}_\m
- \p_\m \vec{A}_\l) - a^{-1} \p_\m \p_\l \vec{A}_\l
+ \vec{v} \times (\vec{A}_\m \times \vec{v})   \cr
& + (v - M) \hat{n} \times \p_\m \vec{\phi}'  + {\rm nonlinear}
}}
\eqn\linearp{\eqalign{
D_5 \vec{\phi} + {{\d S } \over {\d \vec{\phi}}} =
\p_5 \vec{\phi}' &  - \p_\m^2 \vec{\phi}'
+ 2 \l \vec{v} \ \vec{v} \cdot \vec{\phi}'
+ M (\hat{n} \times \vec{\phi}') \times \vec{v}    \cr
& + (a^{-1} - 1) \p_\l \vec{A}_\l \times \vec{v} + {\rm nonlinear},
}}
where have written explicitly only terms that are linear in $A_\m$
and $\phi'$.  These expressions are substituted into the action
$I'_{\rm red}$, eq.\higgsract. Note first that $I'_{\rm red}$ is
quadratic in $\vec{A}_\m$ and $\vec{\phi}'$ (plus higher order terms),
so the classical vacuum is indeed given by $\vec{A}_\m = 0$
and $\vec{\phi}' = 0$.  This corresponds to the classical vacuum of
$\vec{\phi}$ being given by $\langle \vec{\phi} \rangle = \vec{v} = v
\hat{n}$, where the direction $\hat{n}$ is the gauge parameter
introduced in \higgsgf.  The direction of the vacuum in the Higgs phase
is determined by the gauge-fixing.

	The last term in \lineara\ and \linearp\ causes mixing of the
would-be Goldstone boson with the longitudinal part of the $A_\m$
field.  However for the special gauge defined by $M = v$ and $a = 1$ both
mixing terms vanish.  In this case the free
propagators are given by
\eqn\propa{\eqalign{
D^{AA}_{\m\n} & = (1/2)\d_{\m \n} \{ P_\pm
[\o^2 + (k^2 + v^2)^2]^{-1}
+ P_0
[\o^2 + (k^2)^2]^{-1} \}, \cr
D^{\phi \phi} & = (1/2) \{ P_\pm
[\o^2 + (k^2 + v^2)^2]^{-1}
+ P_0 [\o^2 + (k^2 + 2 \l v^2)^2]^{-1} \} \cr
D^{A\phi} & = 0,
}}
where the charged and neutral projectors are
$P_\pm^{bc} = \d^{bc} - \hat{n}^b\hat{n}^c$ and
$P_0^{bc} = \hat{n}^b\hat{n}^c$.
The 4-dimensional propagators are obtained by setting the times equal,
\eqn\fourd{\eqalign{
D^{(4)}(k) = D(t, k)|_{t = 0} = (2\pi)^{-1} \int d\o D(\o, k).
}}
This gives (in the special gauge $M = v$ and $a = 1$),
\eqn\propafo{\eqalign{
D^{(4)AA}_{\m\n} & = \d_{\m \n}  \{  P_\pm
(k^2 + v^2)^{-1}
+ P_0 (k^2)^{-1} \} \cr
D^{(4)\phi \phi} & = P_\pm (k^2 + v^2)^{-1}
+ P_0 (k^2 + 2 \l v^2)^{-1}   \cr
D^{(4)A\phi} & = 0.
}}
One recognizes the free propagators of the 't~Hooft--Feynman gauge in
the 4-dimensional formulation.  In particular only the neutral gauge
particle remains massless, corresponding to the unbroken U(1) symmetry,
and the would-be Goldstone boson, namely the charged components of
$\phi'$, acquires a mass $v^2$. Thus, at the semi-classical level things
behave rather like in the 4-dimensional formulation in this gauge, even
though, as we have seen, at the non-perturbative level the 5-dimensional
formulation of the Higgs phase is quite different from the 4-dimensional
Faddeev-Popov formulation because of the additional condition~\higgsfp.

	For other values of the gauge parameters there is mixing of the
would-be
Goldstone bosons with the longitudinal part of $A$. Indeed the
$A\phi$-term in the quadratic part of $I'_{\rm red}$, eq. \higgsract, is
given, after integration by parts, by
\eqn\crossaf{\eqalign{
I_0^{A\phi} = - (1/2) \int dt d^4x \
& (\hat{n} \times \p_\l \vec{A}_\l)     \cr
& \cdot \Big[ (M - a^{-1}v) \p_5 \vec{\phi}'
+ (v - a^{-1}M)(-\p_\l^2 + v^2)\vec{\phi}'\Big].
}}
The first term vanishes for $M = a^{-1}v$, and the second for
$M = av$, but both vanish only for $M = v$ and $a = 1$.
However one does recover the familiar 4-dimensional free propagators for
the physical degrees	of freedom, namely the transverse $A$-propagator and
the neutral $\phi$-propagator.

   We also give the form of the free propagators for more general values of
the gauge parameters.  For  $M = v$ and $a \neq 1$, the propagators  for
the gauge fields and Higgs fields are:
\eqn\propb{\eqalign{
D^{AA}_{\m\n} = & \ (1/2) P^{\rm tr}_{\m \n} \{ P_\pm
[\o^2 + (k^2 + v^2)^2]^{-1}
+ P_0 [\o^2 + (k^2)^2]^{-1} \} \cr
& + (1/2) P^{\rm lo}_{\m \n} \{ P_\pm [\o^2 + (k^2 + v^2)(a^{-2}k^2 +
v^2)]^{-1} +  P_0 [\o^2 + a^{-2}(k^2)^2]^{-1}  \},
}}
where
$P^{\rm tr}_{\m \n} \equiv \d_{\m \n} - \hat{k}_\m \hat{k}_\n$
and
$P^{\rm lo}_{\m \n} \equiv \hat{k}_\m \hat{k}_\n$, and
\eqn\propc{\eqalign{
D^{\phi \phi} =  \ (1/2) \{ P_\pm
[\o^2 + (k^2 + v^2)^2]^{-1}
+ P_0 [\o^2 + (k^2 + 2 \l v^2)^2]^{-1} \}
}}
\eqn\propd{\eqalign{
D^{\phi^a A_\l^b} = & \ (1/2)(a^{-1} -1) \  \epsilon^{abc} v^c \ ik_\l
\cr & \times [-i\o + k^2 + v^2]^{-1}
[\o^2 + (k^2 + v^2)(a^{-1}k^2 + v^2)]^{-1}.
}}
This gives
$D^{(4)}(k) = [(k^2 + v^2)(a^{-2}k^2 + v^2)]^{-1/2}$
for the charged, longitudinal free $A$-propagator in 4-dimensions, which
does not correspond to any known 4-dimensional gauge.

	Another simplifying gauge choice is $M \neq v$, and $a = 1$, for
which one
obtains
\eqn\prope{\eqalign{
D^{AA}_{\m\n} = & \ (1/2) \d_{\m \n} \{ P_\pm
[\o^2 + (k^2 + v^2)^2]^{-1}
+ P_0 [\o^2 + (k^2)^2]^{-1} \}
}}
\eqn\propf{\eqalign{
D^{\phi \phi} = & \ (1/2) \{ P_\pm
[\o^2 + (k^2 + v^2)(k^2 + M^2)]^{-1}
+ P_0 [\o^2 + (k^2 + 2 \l v^2)^2]^{-1} \}.
}}
\eqn\propg{\eqalign{
D^{A_\l^b\phi^c} = & (1/2)(v - M) \ \epsilon^{bca}\hat{n}^a \ ik_\l \cr
& \times [-i\o + k^2 + v^2]^{-1}[\o^2 +(k^2 + v^2)(k^2 + M^2)]^{-1}
}}
This gives
$D^{(4)}(k) = [(k^2 + v^2)(k^2 + M^2)]^{-1/2}$
for the 4-dimensional propagator of the would-be Goldstone boson, which
again does not correspond to any known 4-dimensional gauge.

\footatend\vfill\supereject\immediate\closeout\rfile\writestoppt
\baselineskip=14pt\centerline{{\bf References}}\bigskip{\frenchspacing%
\parindent=20pt\escapechar=` \input refs.tmp\vfill\eject}\nonfrenchspacing


\bye